# Constructing and explaining machine learning models for chemistry: example of the exploration and design of boron-based Lewis acids

Juliette Fenogli*, Laurence Grimaud*, Rodolphe Vuilleumier*[a]

[a] CPCV, Département de chimie
École Normale Supérieure, PSL University, Sorbonne Université, CNRS
75005, Paris, France

E-mail: juliette.fenogli@ens.psl.eu, laurence.grimaud@ens.psl.eu, rodolphe.vuilleumier@ens.psl.eu

**Abstract:** The integration of machine learning (ML) into chemistry offers transformative potential in the design of molecules with targeted properties. However, the focus has often been on creating highly efficient predictive models, sometimes at the expense of interpretability. In this study, we leverage explainable AI techniques to explore the rational design of boron-based Lewis acids, that activate a wide range of organic reactions. Using Fluoride Ion Affinity as a proxy for Lewis acidity, we developed interpretable ML models based on chemically meaningful descriptors, including *ab initio* computed features and substituent-based parameters derived from the Hammett linear free-energy relationship. By constraining the chemical space to well-defined molecular scaffolds, we achieved highly accurate predictions (mean absolute error < 6 $kJ.mol^{-1}$), surpassing conventional black-box deep learning models in low-data regimes. Interpretability analyses of the models shed light on the origin of Lewis acidity in these compounds and identified actionable levers to modulate it through the nature and positioning of substituents on the molecular scaffold. This work bridges ML and chemist's way of thinking, demonstrating how explainable models can inspire molecular design and enhance scientific understanding of chemical reactivity.

## Introduction

Machine learning (ML) has become indispensable across various scientific domains for uncovering patterns in databases and modeling complex relationships among variables in high-dimensional spaces.[1–3] In chemistry, ML has made significant strides in recent years, leveraging experimental and computed data made accessible by the development of extensive databases,[4,5] high-throughput experiments,[6,7] and super-computers.[8,9] While data-driven modeling and statistical analysis have a long-standing history in chemistry, building upon foundational concepts like the Hammett correlations in 1937,[10] the adoption of ML techniques marks a notable advancement.[11] Indeed, chemical questions ranging from drug discovery,[12] molecular simulations,[13,14] and chemical reaction outcome prediction[5,7,8,15–17] to synthesis planning[18–20] have been successfully addressed.

However, many models, particularly those based on deep neural networks, lack interpretability and are often referred to as "black-box" models,[21] which poses challenges in trust-critical fields like medical diagnosis.[22] Interpretability not only aids in rationalizing decision-making processes but also helps evaluate the validity of a model's predictions against domain knowledge, guarding against spurious reasoning.[23] Furthermore, while some models can be highly accurate, their lack of interpretability limits the ability to extract scientific knowledge from them, thus diminishing their overall interest. The emergence of explainable artificial intelligence (XAI) seeks to address these issues by elucidating what ML algorithms have learned, fostering scientific knowledge, and inspiring new concepts and ideas.[24–26] In chemistry, XAI techniques have been successfully applied to the study of quantitative structure-activity relationships (QSAR),[27–30] drug discovery,[31] and the modeling of organic reactions.[32,33]

Here we investigate how ML can guide the design of molecules with a targeted specific property: the Lewis acidity, which is of high importance in a wide range of organic transformations.[34,35]

Therefore, predicting the Lewis acidity of compounds is highly valuable and has recently been explored by the Greb group.[36] An advanced graph neural network (GNN) model was trained on a substantial database of nearly 49k Lewis acids with diverse central atoms and ligands to predict the Lewis acidity of any compound. Complementarily, we aim at expanding the exploration of Lewis acidity using XAI, focusing on restricted chemical spaces, employing "white-box" self-explanatory models or XAI *post-hoc* techniques. Interpreting ML models can enhance chemical understanding by acting as a "computational microscope",[24,25] to explore the qualitative concept of Lewis acidity and as a "source of inspiration"[24,25] for designing novel boron compounds with targeted Lewis acidity. Strategies to boost the Lewis acidity of borane derivatives include introducing electron-withdrawing groups onto ligands[37–40] or using constrained geometry ligands.[41–43] While functional group decoration is common in QSAR organic chemistry studies, applying ML techniques allows for the exploration of a much broader chemical space, providing novel insights for molecule design.

In this study, we employed ML techniques to create and analyze arrays of Lewis acids, annotated with their corresponding Lewis acidity, measured via the fluoride ion affinity (FIA) (Figure 1). To improve explainability, we restricted the explored chemical space to four boron derivative scaffolds possessing aromatic rings substituted by electron-donating or -withdrawing substituents. Combining several molecular descriptors and ML algorithms, we built regression models to predict FIA and interpreted them to decipher the intrinsic Lewis acidity of these compounds and identify actionable levers for molecular design.

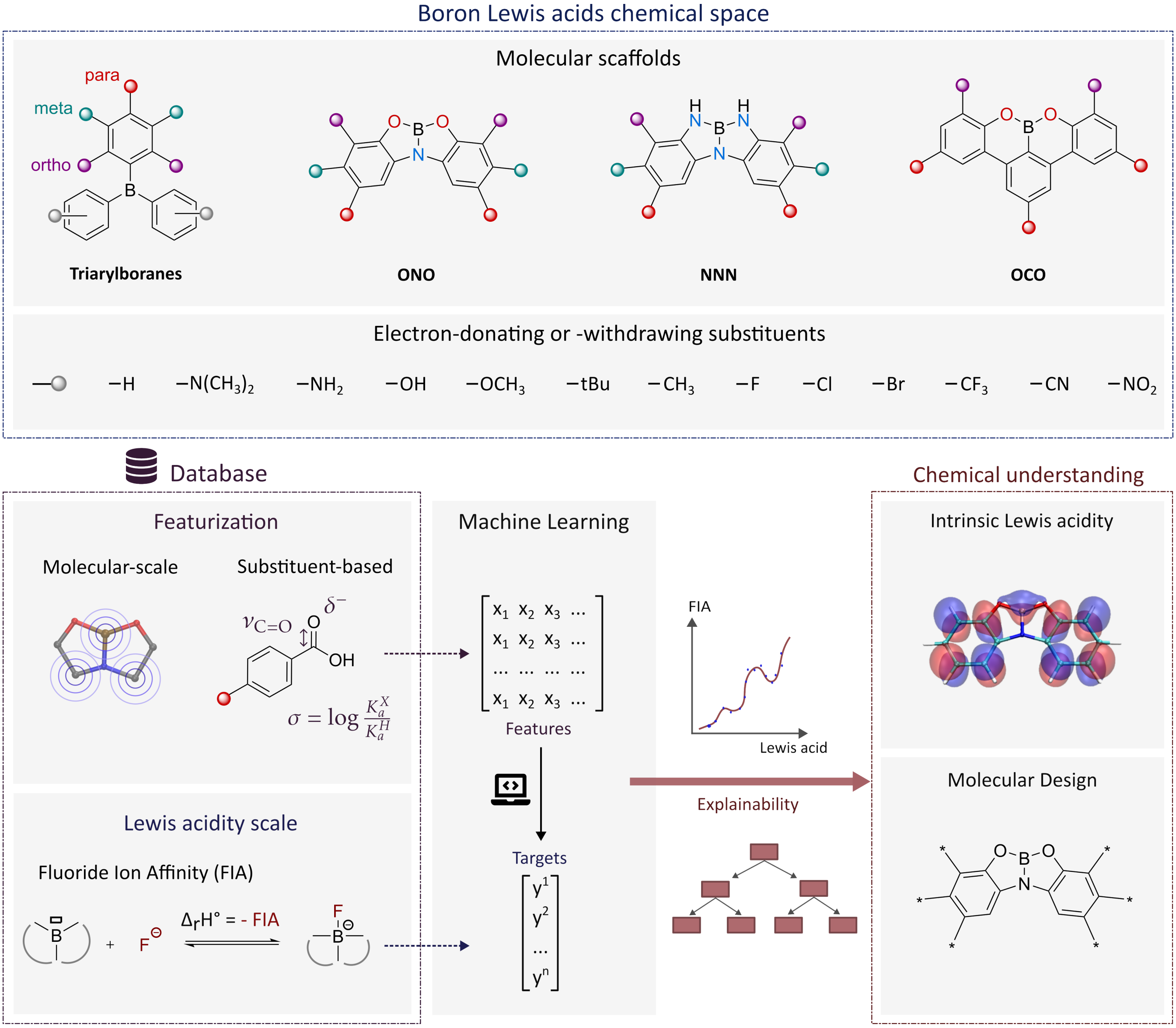

**Figure 1.** Workflow towards the design of efficient boron Lewis acids: spanning from defining chemical space (four molecular scaffolds and substituents) and molecule features (chemo-informatics, *ab initio* computed or derived from the Hammett correlations) to labeling Lewis acidity by FIA and constructing predictive models. Models' predictions were then explained to get insight into the qualitative concept of Lewis acidity and design boron Lewis acids with specific Lewis acidity profiles.

# Results and Discussion

## Lewis acidity scale

To construct a database of boron Lewis acids, a robust metric for quantifying Lewis acidity was essential. The chosen metric needed to be readily accessible to facilitate the fast labeling of molecules, while also ensuring consistency, as noisy datasets can hinder the learning process and potentially lead to overfitting.[44] Several metrics were evaluated and benchmarked with conventional scales employed in experimental organic chemistry to determine Lewis acidity (see the SI, *section S1*). FIA, representing the standard negative enthalpy change associated with the binding of a fluoride ion to the Lewis acid in the gas phase (Figure 1), was retained as the most relevant, consistent and accessible quantity, as it is a computed metric. Various cost-effective *ab initio* methods were evaluated to compute FIA, ensuring alignment with values computed at a higher level of theory[45] (SI, *section S2*). We used density functional theory (DFT) at the M062X/6-31G(d) level of theory in isodesmic calculations as a compromise between efficiency and precision to provide reliable FIA data. Throughout this manuscript, FIA has been used as a proxy for Lewis acidity. Therefore, any conclusions drawn about FIA also apply to Lewis acidity, and the terms have been used equivalently.

## Chemical space

We targeted scaffolds possessing aromatic ligands, which are particularly effective at transmitting electronic effects over long distances through their electronic π systems, unlike aliphatic or non-conjugated ligands. This facilitates the combination of effects from multiple substituents on the same central atom. Additionally, as aromatic positions are not chemically equivalent, their diverse substitution allows to precisely modulate the effect on FIA. While the substitution of aromatic positions to modulate molecular reactivity is well-established in organic chemistry,[38] its systematic integration with ML to predict Lewis acidity is, to our knowledge, unprecedented. Previous databases, such as the one developed by Greb and coworkers,[36] have primarily focused on exploring the effects of the central atom type and ligand denticity. To enhance model performance and simplify interpretation, we limited the chemical space to four scaffolds: triarylboranes, which have been extensively studied for their catalytic activity,[34,35] and three constrained Lewis acid scaffolds (ONO[46], NNN[42], OCO[47]) featuring heteroatoms coordinated to boron. The planar geometry of these constrained ligands minimizes steric hindrance, while the pincer-like structure facilitates boron atom accessibility, enhancing Lewis acidity.[42] Each dataset consists of symmetric molecules, substituted with 13 possible electron-donating or -withdrawing substituents at specific aromatic positions, to minimize steric hindrance and ensure synthetic efficiency (Figure 1Error: Reference source not found). These constraints result in an accessible chemical space of 2197 molecules for the ONO scaffold for example (Figure 2.A). Initially, datasets were generated through random substitution, prioritizing hydrogen atoms over other functional groups. This approach was intended to primarily examine the impact of having only one or two non-hydrogen substituents but led to an overrepresentation of such molecules (clusters in Figure 2.A, ONO dataset). To enhance diversity and better represent the chemical space for model development, the ONO dataset was expanded from 175 to 272 molecules. This expansion used k-means[48] and coverage[49] algorithms applied to molecular fingerprint representations, to uniformly sample the ONO chemical space (Figure 2.A, additional details in SI, *section S3*). Datasets of the OCO, NNN and triarylboranes scaffolds contain less molecules (61, 80 and 181, respectively) and were mainly used to assess the extrapolation capabilities of the developed models and to compare the effects of substituents on Lewis acidity across different scaffolds for molecular design.

Molecules in the four datasets were labelled by their computed FIA value. The range of FIA values varies depending on the ligand scaffold (Figure 2Error: Reference source not found.B). OCO scaffold exhibits a very narrow distribution of FIAs, whereas FIA is relatively normally distributed across ONO derivatives, which will be advantageous when constructing ML models. In contrast, triarylboranes show a more uniform distribution, offering a broader spectrum of accessible FIA values. Among these, the strongest Lewis acids are found in the ONO and triarylboranes scaffolds.

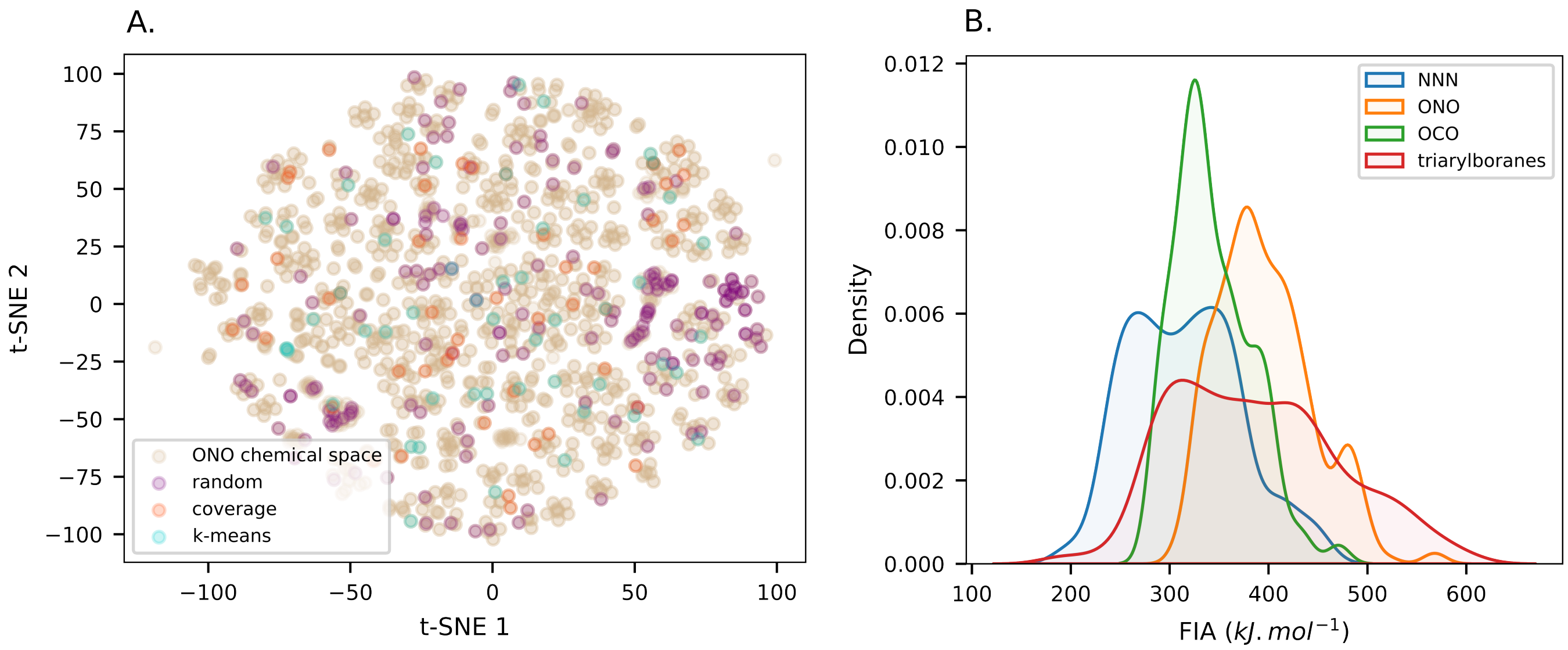

**Figure 2.** Studied chemical space. A. t-SNE map on fingerprint representation showing the ONO chemical space (2197 molecules, light brown circles), the initial random dataset (purple circles) and the dataset extension thanks to k-means (light blue circles) and coverage (light orange circles) algorithms (see SI, *section S3)*. B. Kernel density estimation plot of the FIA distributions for each molecular scaffold.

## Constructing models

Creating performant models is a prerequisite to any interpretability task to ensure the reliability of the interpretations. Since we aimed to design novel boron Lewis acids, we focused on the ONO molecular scaffold, which we had designed in the lab, for building ML models. A benchmark of ML models and molecular descriptors was realized to identify the most effective combinations to predict FIA (Figure 3). Models were optimized on a training set following a grid search algorithm and 10-fold cross-validation (CV). They were subsequently evaluated on a testing set (see the SI, *section S5*, for details on dataset splitting). Among the evaluated molecular descriptors, some were already implemented in Python libraries (like RDKit[50] or DeepChem[51]), and are commonly used in drug design, such as Morgan fingerprints[52,53] and RDKit descriptors. While Morgan fingerprints are beneficial for visualizing fragment diversity within chemical space (Figure 2.A), they are less effective at accurately predicting FIA, with a mean absolute error (MAE) higher than 10 kJ.mol$^{-1}$. This limitation arises because they focus primarily on local connectivity, probably missing the effect of the delocalized π electrons on aromatic rings. RDKit descriptors yield highly effective models, such as linear models or SVR (Figure 3.Error: Reference source not foundA), yet out of the 208 features generated (e.g., counts of fragments, partial charge or molecular weight), only a few dozen is readily interpretable. More physics-based parameters can be obtained through the Auto-QChem[9] DFT calculation workflow yielding DFT-derived molecular (e.g., frontier orbitals energies) and atomic descriptors for the boron atom (e.g., partial charge), totaling 43 features. These descriptors, referred to as “quantum descriptors”, offer reasonably good performance across all ML models but require significantly more computational resources. Despite of this calculation cost, these descriptors offer reliable insights into the relationship between the quantum features of molecules and their Lewis acidity (see part **Interpretability** – *insights in the Lewis acidity*). However, quantum parameters are not directly manipulable by organic chemists for designing molecules. Instead, the typical strategy involves designing molecular structures by substituting various functional groups on a molecular scaffold, which play a significant role in modulating Lewis acidity. To better understand this influence, particularly the electronic effects of substituents, we introduced substituent-based molecular descriptors. This method builds upon the foundational work of Hammett,[10] who derived the substituent constants $\sigma_m$ and $\sigma_p$, for the *meta* and *para* substituents respectively, establishing the groundwork for what would become a pioneering approach to QSAR. However, Hammett constants are limited to *meta* and *para* substituents on aromatic rings, as *ortho* substituents complicate the analysis by introducing both electronic and steric effects.[54] Indeed, when considering only *meta*- and *para*-substituted molecules from the triarylboranes dataset, the trend of FIA follows a linear relationship with $\sigma_m$ and $\sigma_p$ (FIA = 243 $\sigma_m$ + 91 $\sigma_p$ + 351, determination coefficient $R^2$ = 0.91) (see Figure S9.A). Conversely, when considering also the triarylboranes possessing a substituent at the *ortho* position, the linear relationship is compromised (Figure S9.B, $R^2$ = 0.63), and this effect is intensified when considering other molecular scaffolds such as ONO ($R^2$ = 0.54), for which the labeling of *ortho*, *meta* and *para* substituents is less clear. To address this issue, we implemented "Hammett-extended descriptors", as proposed by Sigman et al.[54] Additionally to $\sigma_m$ and $\sigma_p$, these descriptors encompass computationally derived steric and electronic parameters for substituents across all three aryl positions (*ortho*, *meta*, and *para*) of benzoic acid. They include infrared (IR) carbonyl stretching ($\nu_{C=O}$) and COH bending ($\nu_{COH}$) frequencies and intensities, natural bond orbital (NBO) charges of each atom in the carboxylic acid moiety, Sterimol $B_1$, $B_5$, and $L$ of the substituent,[55] and the torsion angle between the carbonyl group and the aromatic ring plane when the substituent is positioned in *ortho*. However, these descriptors have limited applicability because they can only be applied to one molecular scaffold at a time, since they do not describe the carbon backbone of the molecule (Figure S10). For the ONO scaffold, the *ortho*, *meta*, and *para* positions were conventionally defined relative to the oxygen bonding atoms. This convention was extended to the NNN scaffold, but these descriptors were not implemented for the OCO. Using SMARTS substructure identifiers[56] implemented in the RDKit Python library,[50] the chemical nature of substituents at *ortho*, *meta*, and *para* positions was identified. Then Hammett-extended parameters derived by Sigman and co-workers[54] corresponding to the *ortho*, *meta* and *para* substituents were concatenated into a vector featuring the molecule (36 features). These descriptors demonstrate robust performance with advanced models such as gradient boosting (Grad. Boost.) and multi-layer perceptron (MLP) regressors. They also excel with simpler models like linear and linear ridge (LR) regressors, achieving a mean absolute error of approximately 6 $kJ.mol^{-1}$ (Figure 3.C). Given the straightforward nature of the molecule description and the impressive results obtained with these basic models, they provide an optimal balance between interpretability and prediction accuracy. These findings align with Hammett's theory of linear free energy relationships (LFER), which suggests that functional groups on an aromatic ring near the reactive site linearly influence the molecule's reactivity.

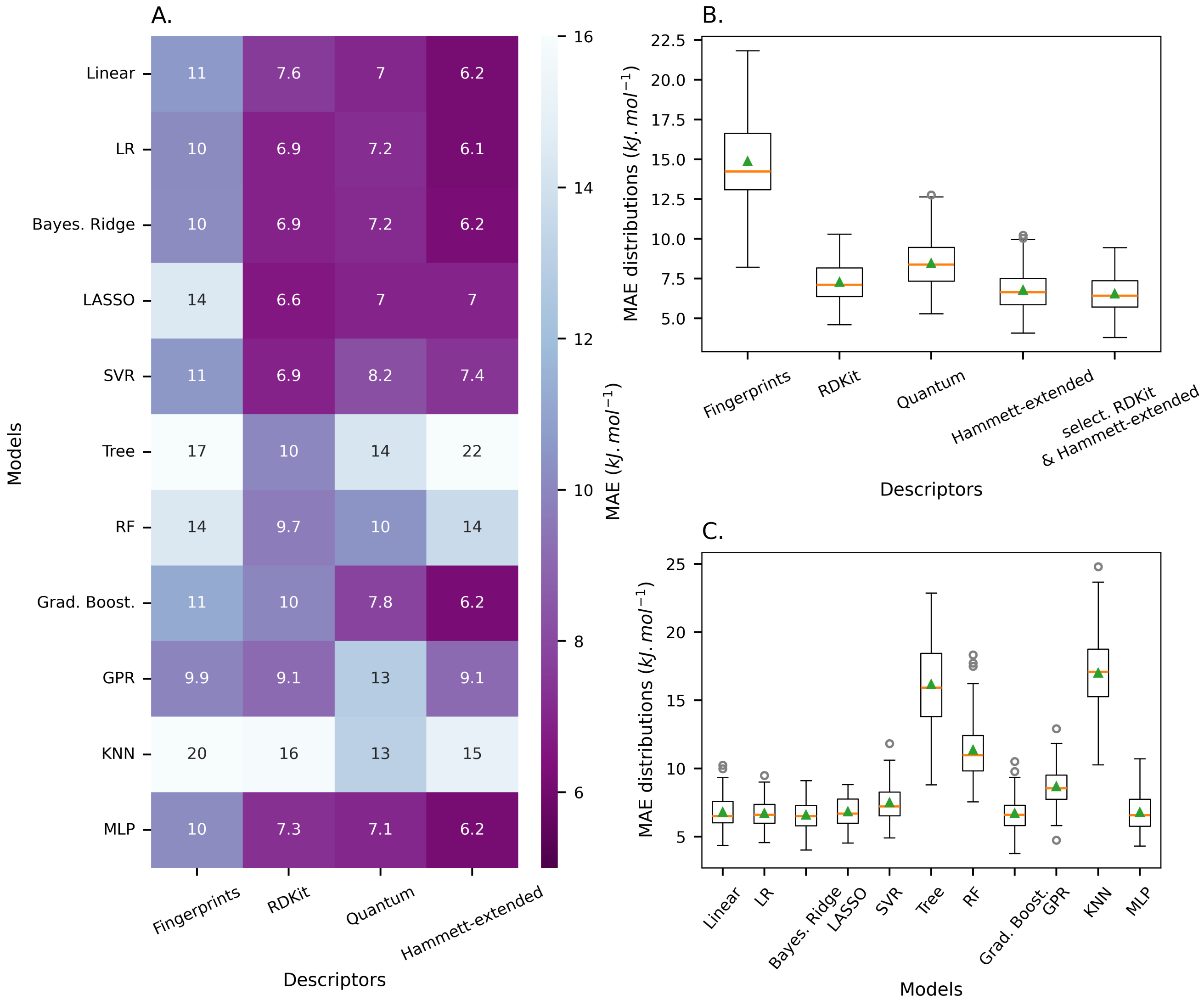

**Figure 3.** Benchmarking machine learning models and descriptors in FIA prediction. A. Heatmap of MAE scores calculated on the testing set for different combinations of models and descriptors (Linear: linear regression, LR: linear ridge, Bayes. Ridge: Bayesian ridge, LASSO: least absolute shrinkage and selection operator, SVR: support vector regression, Tree: decision tree, RF: random forest, Grad. Boost.: gradient boosting, GPR: gaussian process regressor, KNN: K nearest neighbors, MLP: multilayer perceptron), B. Boxplots of MAE obtained on the training set for the linear ridge (LR) model, combined with different molecular descriptors C. Boxplots of MAE scores on the training set for the Hammett-extended descriptors, combined with different regressors (10-fold CV repeated 10 times with different folds).

From this benchmark (Figure 3Error: Reference source not found.A), we recognized the importance of both global descriptors and substituent-based descriptors in predicting FIA, which is why RDKit and Hammett-extended descriptors were concatenated in an attempt to create a high-performant predictive model, an oracle. Apart from the challenge of creating an accurate predictive model, having such a model would be helpful to investigate molecular design (see part **Interpretability – *Molecular design*)*Chemometrics on ONO chemical space*. For any given molecule, one could quickly determine its precise FIA, as these concatenated descriptors are also fast to compute. This yielded 244 features, from which we selected 126 based on their correlation with the target FIA, evaluated using the F-statistic. We selected LR as the machine learning algorithm due to its strong performance in predicting FIA from these two molecular descriptors. This allowed to reach a MAE of 6.39 $kJ.mol^{-1}$ on the training set (repeated 10-fold CV, Figure 3Error: Reference source not found.B) and of 5.39 kJ.mol$^{-1}$ on the testing set (Figure 4, $R^2$ = 0.98), representing less than 2% error relative to the average FIA value (around 450 kJ.mol$^{-1}$). The performance on the ONO dataset, with a MAE between 5 and 10 kJ.mol$^{-1}$ across almost all ML algorithms and molecular descriptors (Figure 3.A), can be attributed to the limited chemical space it encompasses. In contrast, the optimized graph neural network model of Greb and co-workers, evaluated on a wide variety of molecular structures, struggles to achieve a MAE below 10 kJ.mol$^{-1}$. This model gave a MAE of 23 kJ.mol$^{-1}$ ($R^2$ = 0.51) evaluated on the testing dataset of ONO scaffold (in comparison with the MAE of 5.39 kJ.mol$^{-1}$obtained by the present oracle). In addition, such deep learning (DL) models necessitate huge amounts of data to perform and are not designed for tasks in low-data regime.

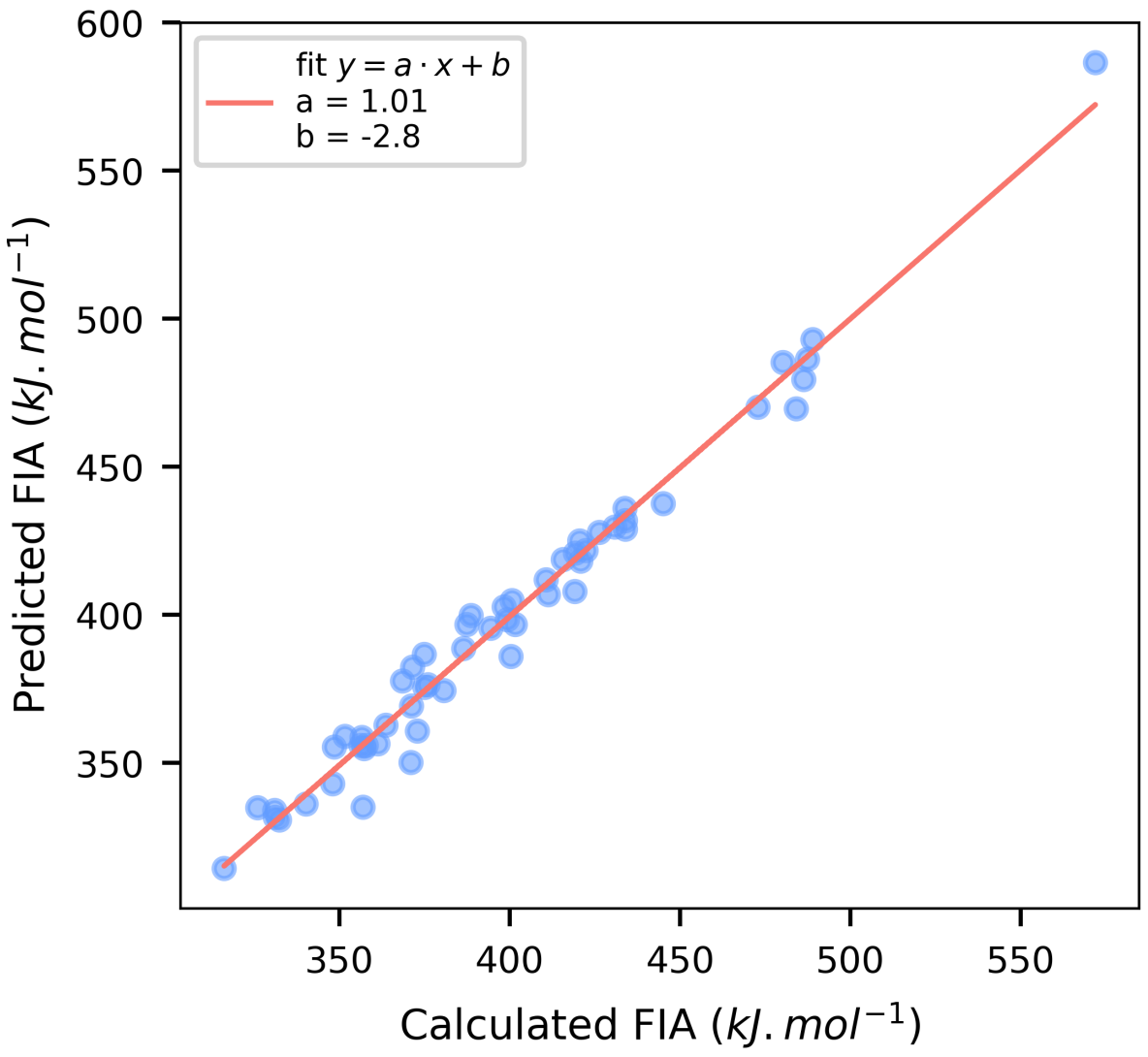

**Figure 4.** Optimized model for ONO evaluated on the testing set (MAE = 5.39 kJ.mol$^{-1}$).

## Capacity to extrapolate to another molecular scaffold

After optimizing models to predict FIA within the restricted ONO chemical space, we aimed to assess their ability to predict FIA across different chemical spaces, particularly from a molecular scaffold to the other. We also explored methods, such as feature selection, to customize models for enhanced versatility. However, not all descriptors are suitable for this purpose. In addition to being difficult to interpret, fingerprints, represented as bit-vectors, have interdependent features that cannot be individually selected. Furthermore, Hammett-extended descriptors rely solely on the chemical properties of substituents. Molecules with identical *ortho*, *para*, and *meta* substituents are then assigned the same molecular descriptors, even though their FIAs can differ significantly across structures (refer to Figure S10). For this reason, we focused on enhancing the extrapolation capabilities of models based on quantum or RDKit descriptors (see the SI, *section S5* for the latter).

Models trained on ONO dataset, featured by the quantum descriptors, can generally predict the FIA trend for the NNN, though with a model-dependent bias that impacts the MAE (Table S7). Simple models struggle to extrapolate, whereas the Multilayer Perceptron (MLP) achieves reasonable prediction performance (MAE = 24.1 kJ.mol$^{-1}$), as expected given the inherent versatility of neural networks. This is illustrated in Figure 5.A, showing the prediction for the NNN from a LR model trained on ONO. While the trend is well captured (Pearson's r = 0.96), the MAE is high, reaching 188 kJ.mol$^{-1}$. The bias in prediction observed in most models can be partly attributed to the considerable variation in some features ranges between the two molecular scaffolds. When the model heavily relies on such features, predictions from one structure to another can be impacted by the high variation, leading the model to overestimate the features' effect. Even if these features are important for initial predictions, removing them may improve extrapolation. Additionally, features not correlated with the target FIA can be removed to streamline the models and enhance their generalizability.

Therefore, we ranked features based on their differences between ONO and NNN structures (Table S9) and their correlation to the target FIA (Table S10). We assessed the prediction performance of the LR model by systematically removing features using these two criteria. Among the features not correlated with FIA, boron atom parameters such as its coordinates and buried volume led to a notable drop in MAE (from 188 to 158 $kJ.mol^{-1}$) when removed. Eliminating the lower unoccupied molecular orbital (LUMO) energy, that is strongly correlated with FIA (Pearson's r = 0.628) but varies considerably between structures, decreased the MAE to 140 $kJ.mol^{-1}$. Similarly, although not correlated with FIA (Pearson's r = 0.076), the population of Rydberg atomic orbitals (NPA_Rydberg) changes significantly between the two structures. Removing this feature decreased the MAE to 14.5 $kJ.mol^{-1}$, indicating a considerable contribution to the observed bias, albeit difficult to explain. When this feature alone was excluded, the MAE dropped directly from 227 to 35.5 $kJ.mol^{-1}$. Further removal of the dipolar moment of the molecule reduced the MAE to 13.6 $kJ.mol^{-1}$. The improvement in prediction performance of the LR model with the selected features is illustrated in Figure 5.B. Prediction performance remained high within the ONO (8.26 $kJ.mol^{-1}$) and the NNN (11.4 $kJ.mol^{-1}$) chemical spaces using these selected features. However, it's important to note that the feature selection is task-specific, as evidenced by varying performance in predicting from ONO to OCO (MAE = 84.8 $kJ.mol^{-1}$) and triarylboranes (MAE = 759 $kJ.mol^{-1}$) using these features. This is due to inherent differences between scaffolds, notably in terms of FIA distribution (Figure 2.B) and electronic structure (Figure 6). To explore these differences, we attempted training on three molecular structures while testing on the fourth, which partly improved the extrapolation performances (see the SI, *section S5*).

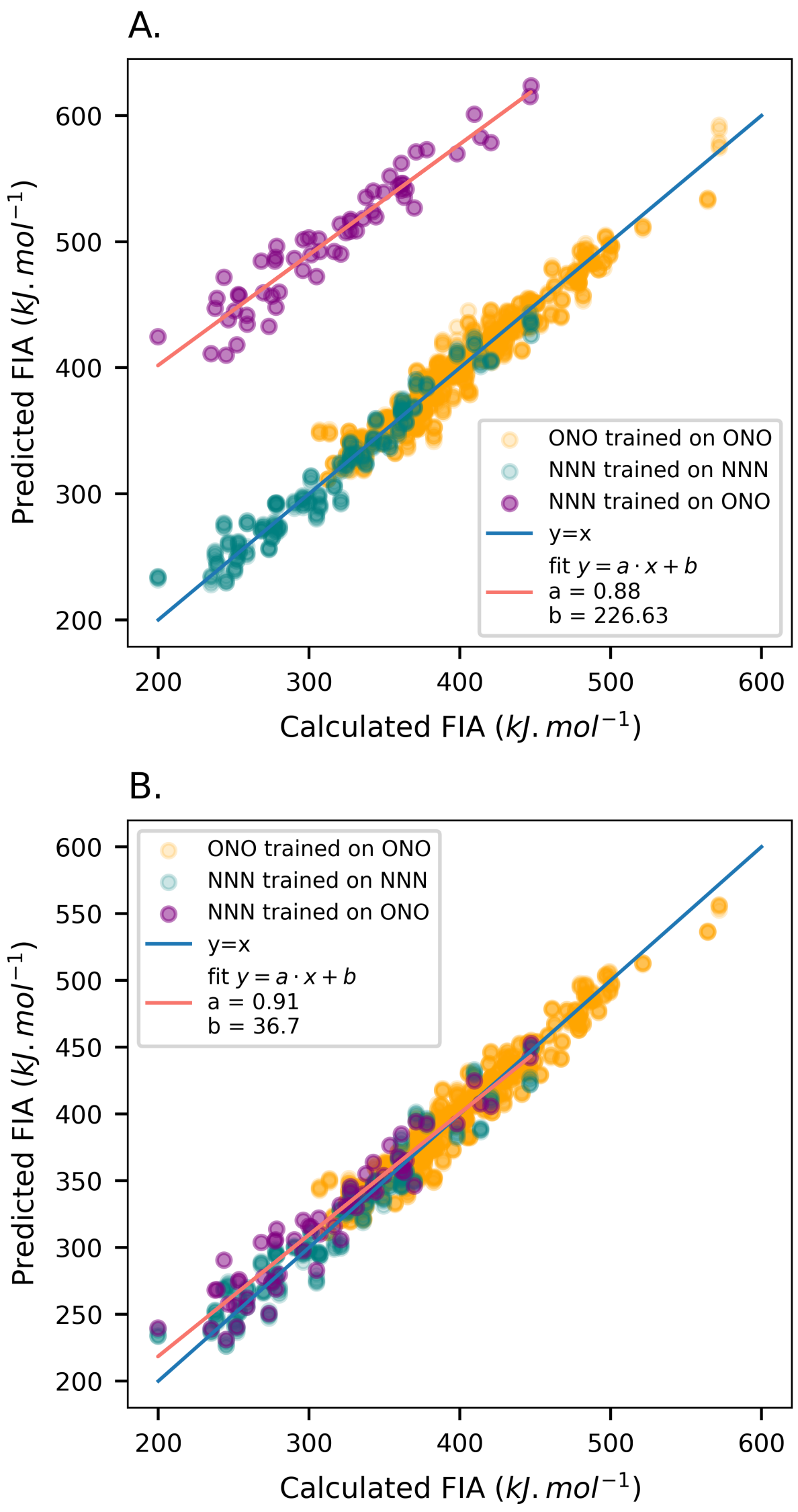

**Figure 5.** Feature selection to extrapolate from ONO to NNN with the LR model and quantum descriptors. A. No feature selection. B. Features selected.

## Interpretability

Our goal here is to address two main questions by interpreting the developed models. First, we aim to gain insight into the intrinsic nature of Lewis acidity for boron derivatives. Second, we seek actionable explanations for a compound's Lewis acidity, exploring new routes to design molecular structures with targeted Lewis acidity.

### *Insights in the Lewis acidity*

ML can serve as an instrument to unveil properties of a system that are challenging or even impossible to probe using traditional methods. Lewis acidity is inherently a qualitative concept. It can be accessed through the reaction enthalpy of a Lewis acid-Lewis base adduct formation, but unlike electrophilicity, that is associated with the energy level of the LUMO, the relationship between molecular properties and Lewis acidity remains elusive. Analyzing FIA predictive models may help to unravel the origins of Lewis acidity at the electronic scale.

We used quantum descriptors, that provide precise physical parameters for characterizing molecules, and examined the whole database (the four molecular scaffolds together) to find broader patterns and insights. Conducting a principal component analysis (PCA) to simplify complex data into two dimensions, we identified distinct groups of molecules (Figure 6). Notably, the triarylboranes appear separate from other structures along the PC1 axis, which primarily encompasses electronic parameters of the boron atom. This separation is unsurprising as triarylboranes lack heteroatoms bonded to boron, which can affect electronic population through electron-withdrawing effects. Additionally, the two PC derived from quantum descriptors reflect FIA evolution across the database. This confirms the relevance of these descriptors in capturing Lewis acidity.

We then simplified the data by choosing twelve significant but uncorrelated features through a hierarchical clustering (SI, *section S6*). Most ML methods struggled with these simplified descriptors across the four molecular scaffolds, except for tree-based ensemble models like Random Forest (RF) and Grad. Boost., which performed better than the others (see Table S17). Our goal was to explore how these features impacted the FIA using two interpretable ML approaches. We first looked at a straightforward and simple model using linear regression to understand how much each feature contributed. Then, we used the permutation feature importance *post-hoc* technique, with a Grad. Boost. regressor, as this model was the most performant on uncorrelated features (MAE = 14.6 $kJ.mol^{-1}$, Table S17). The learned coefficient of each feature in the linear model, as depicted in Figure S13.C, reveals its importance. Both the global electronegativity of the molecule and the partial charge of the boron atom (NPA_charge) emerge as the most significant features, with approximately equal importance. This observation aligns with their ranking in the permutation feature importance, although the electronegativity being slightly more influential (Figure S13.B). When the regression is restricted to the ONO dataset (Figure S15), electronegativity becomes the most significant parameter, with the partial charge coefficient being three times lower. Electronegativity is a linear combination of frontier orbital energies, suggesting that, for the boron derivatives studied, Lewis acidity is somewhat more influenced by molecular orbital interactions than by Coulombic interactions. This is notable because FIA is traditionally considered as an index of hard acidity, where Coulombic interactions would be expected to dominate.[45,57]

The absolute electronegativity is then a key global factor in determining FIA magnitude, while local electronic parameters of boron atom more precisely adjust the predicted FIA. FIA can be roughly linearly predicted using only these two parameters for the ONO scaffold ($FIA = 60.0\ \chi + 8.15\ NPA\ charge + 161$, $R^2$ = 0.88, Figure S16). Steric parameters (e.g., molar_volume) seem to play a less significant role, potentially due to the predominance of constrained-ligand boron derivatives in our database. Thus, controlling the electronic environment around the boron atom is essential. While chemists cannot directly manipulate quantum properties that are not “actionable”, they can adjust Lewis acidity through molecular design by varying functional groups on the molecular scaffold. However, this approach is complicated by the interconnected effects of the substituents on the same reactive site, making the use of ML relevant.

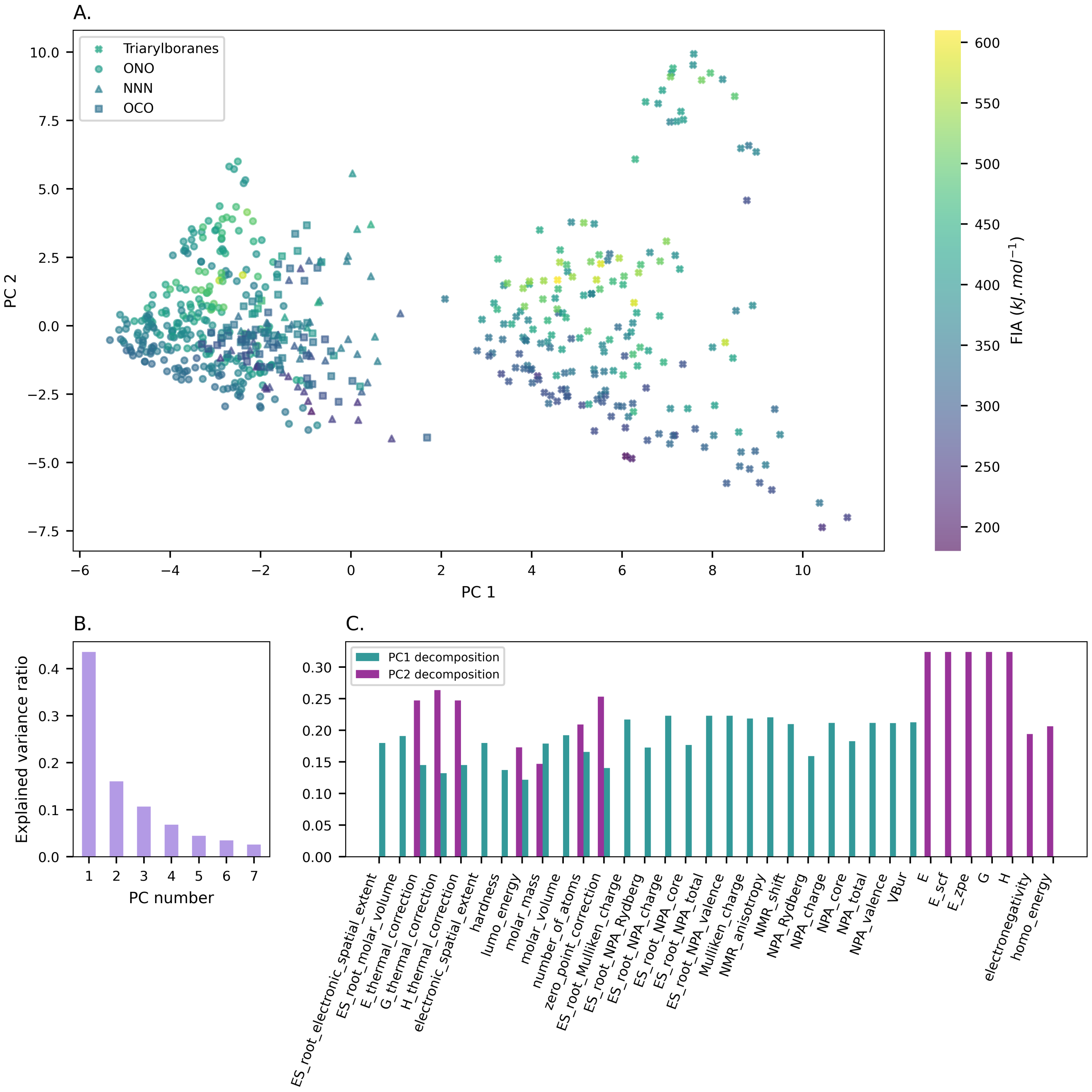

**Figure 6.** Principal component analysis (PCA) of the quantum descriptors. A. Projection onto the first two principal components. Molecules are colored according to their FIA values. Triarylboranes (small crosses) appear clearly separated from other molecular scaffolds in terms of electronic structure of the boron atom. B. Explained variance ratio for each component. C. Contributions of the quantum descriptors features to the first two principal components.

*Molecular design*

<u>Interpretable ML</u>

This part will focus on the ONO molecular scaffold (see *SI*, *section S6* for a comparison with the triarylboranes). We used Hammett-extended descriptors, that are specifically tailored to unravel the effect of substituents.

Before any regression by a ML algorithm, we examined the correlation between *ortho*, *meta*, and *para* parameters, relative to the oxygens, and FIA (Figure S17.A), which shows that *para* substituents have the most significant effect on FIA. Steric parameters, such as the torsional angle $\theta_{tor}$ or Sterimol parameters, poorly correlate with FIA. We then selected twelve uncorrelated features through hierarchical clustering (Figure S18.A). Using again the linear model (Figure S17.B) and a permutation importance algorithm (Figure S18.B) on a Grad. Boost. regressor, we identified the Hammett $\sigma$ constants as the most significant parameters, $\sigma_p$ being slightly more important than $\sigma_m$, with a lower contribution of parameters characterizing the electronic demand of the *ortho* substituent. This observation is consistent with the correlations with FIA (Figure S17.A).

These methods help identify which aromatic positions to prioritize when designing an ONO compound with a specific Lewis acidity, namely here, the *para* position. However, they do not inform on the electronic demand of the substituent or the interactions between substituents at different positions. We used decision trees to understand the decision criteria for predicting FIA. We thus categorized FIA into six classes based on the ONO chemical space distribution (Figure 2.B), turning the original regression task into a classification problem. To simplify the analysis and the resulting tree structure, we focused on four Hammett-extended features capturing steric and electronic effects of *ortho*, *meta*, and *para* substituents. We included the partial charge of the oxygen atom of the C=O bond of benzoic acid ($NBO_{=O}$) for each position, as the Sigman group showed that these parameters could effectively replace the empirical $\sigma$ constants,[54] plus $L_o$, the bond length between the ortho substituent and the aromatic ring, to account for steric effects at the *ortho* position. The root node of the decision tree shows that the $NBO_{=O}$ charge for the *para* substituent is the critical criterion for classifying LA as good, strong or super, if it exceeds -0.59*e*, a threshold distinguishing mesomeric electron-withdrawing groups (such as CN and $NO_2$) from others (see Table S6); otherwise, it is classified as medium. Subsequent nodes evaluate the *ortho* and *meta* substituents for majority-medium and majority-good classes, respectively. The obtained tree diagram is depicted in Scheme 1, translated into chemical terms based on the feature thresholds from the original Scikit-learn[58] tree (see Figure S19).

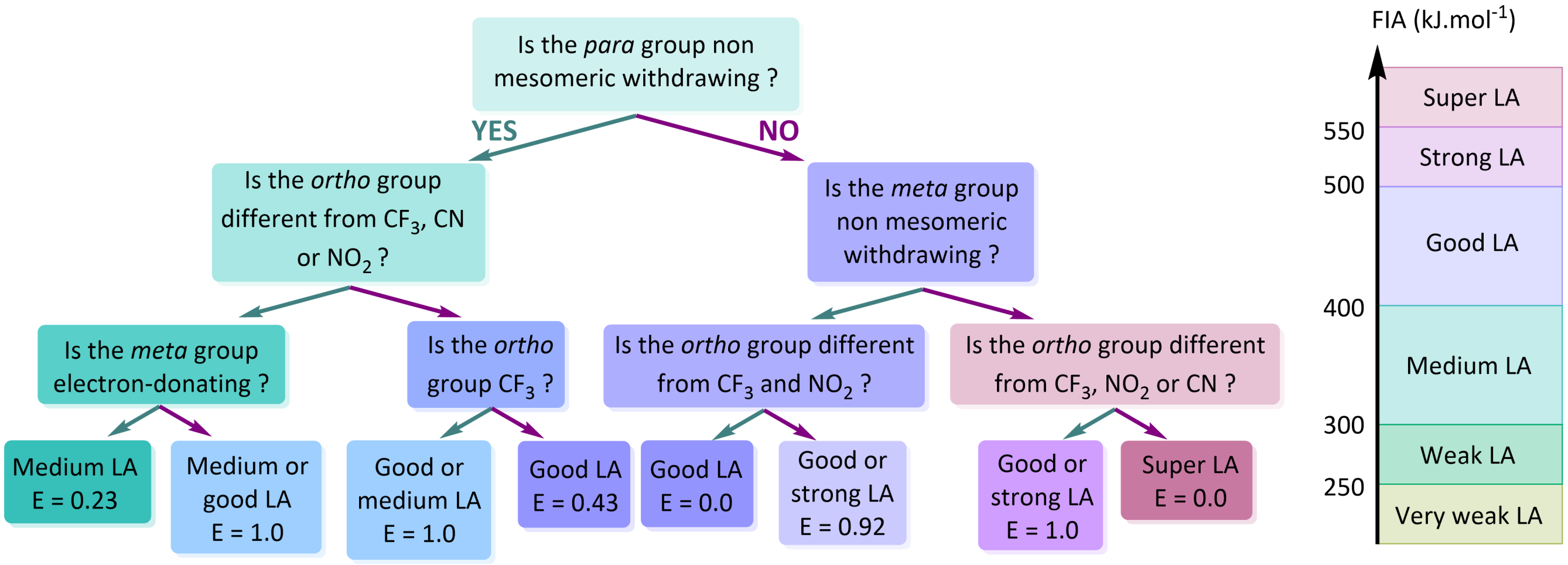

**Scheme 1.** Decision tree model for the ONO structure (0.73 accuracy on 10-fold CV). Nodes are split based on Hammett-extended feature thresholds. A left arrow indicates a “Yes” decision, while a right arrow indicates a “No” decision. The entropy (E) of each node reflects its purity, indicating whether it contains a unique class (E=0) or not. Original tree from Scikit-learn[58] is provided in Figure S19.

In summary, the tree structure shows that adding or not a mesomeric electron-withdrawing group at the *para* position of an ONO molecule already determines the accessible range of FIA (Scheme 1). Then, the substituents at the *ortho* and *meta* positions must be examined to refine the targeted FIA value. To confirm these results, we need more data for statistical analyses on the diverse substituents. For that, we have used the oracle previously developed to screen the entire ONO chemical space and provide precise FIA values to enrich our database.

Chemometrics on ONO chemical space

Using the root node criterion from the tree in Scheme 1, we compared the FIA distributions for ONO molecules with a mesomeric electron-withdrawing group in the *para* position to those without. The 2197-molecules chemical space is effectively separated as shown by the weak overlap between violin plots of FIA distributions (Figure 7.A). Next, we divided the set of molecules possessing a mesomeric electron-withdrawing group in *para* by their *ortho* group (Figure 7.B). Since these distributions are still broad, merely fixing the *ortho* group while maintaining a *para* mesomeric electron-withdrawing group fails to standardize the FIA. Therefore, we put the same substituent at both the *meta* and *ortho* positions, resulting in a single molecule, with a fixed FIA value. The approach ensures that the electronic demand at the *ortho* and *meta* positions is nearly the same, reducing the complexity of the interconnected effects of the substituents. Varying this substituent, while keeping a mesomeric electron-withdrawing group in *para*, enables exploration of the full range of FIA values (from 400 to 600 $kJ.mol^{-1}$) for the ONO scaffold (see the spread of ONO FIA distribution, Figure 2.B).

This analysis suggests that if an organic chemist needs a compound within a specific FIA range, say 450 to 500 $kJ.mol^{-1}$, a practical approach using the ONO scaffold might involve adding a mesomeric electron-withdrawing group in *para* and halogens (such as -F, -Br, or -Cl) in *ortho* and *meta* positions, although synthetic challenges should be considered when implementing this strategy.

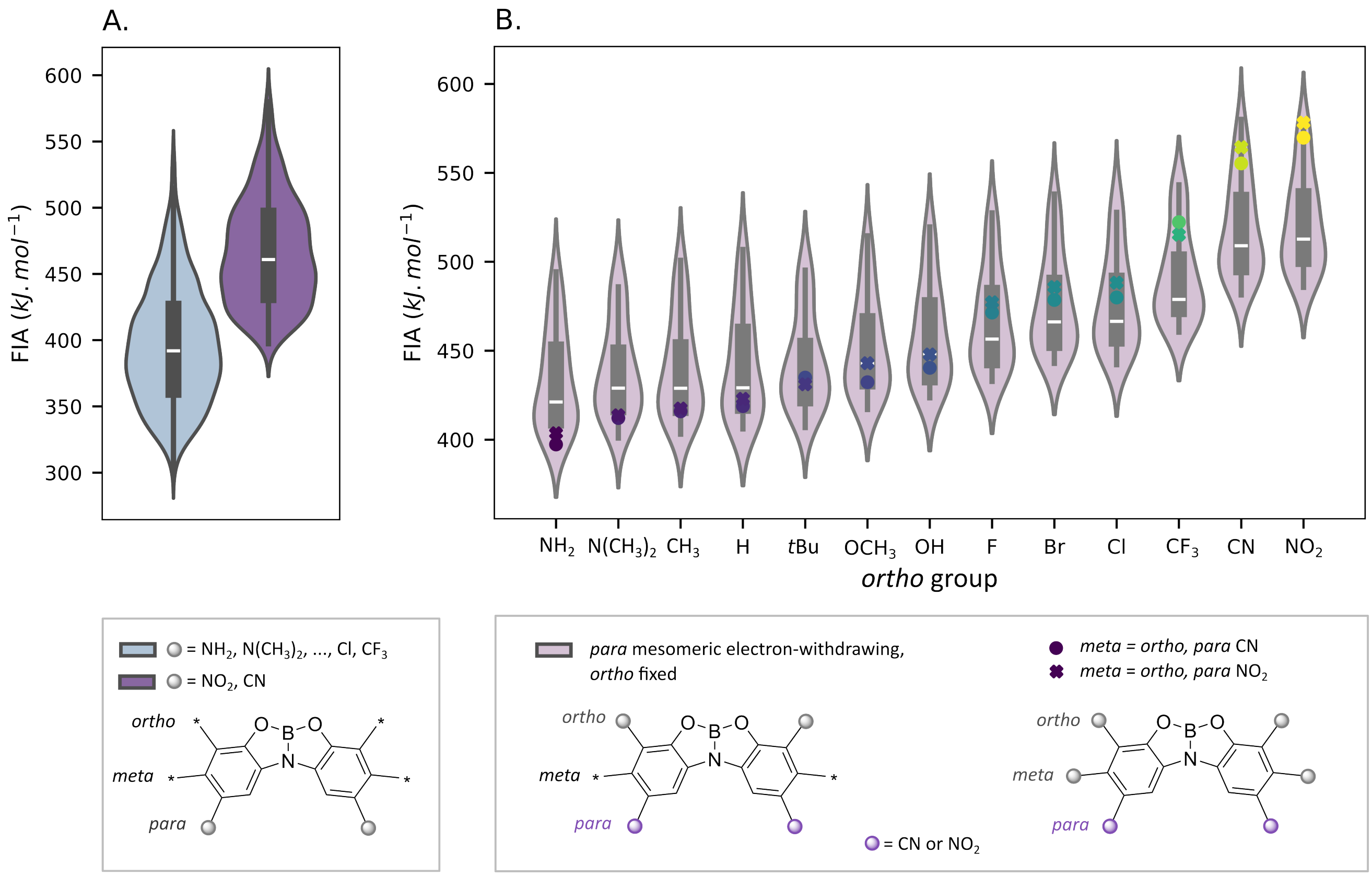

**Figure 7.** A. Violin plots separating data between molecules possessing mesomeric electron-withdrawing groups (CN or $NO_2$) in *para* position or not. B. Distributions of molecules with a mesomeric electron-withdrawing group in *para*, separated by the *ortho* group and highlighting variations in the *meta* group; dotted markers indicate molecules where the *meta* group matches the *ortho* group.

# Conclusion

ML has proven highly effective in predicting Lewis acidity via FIA regression models. Advanced techniques like GNN, trained on extensive datasets of tens of thousands of molecules,[36] achieve MAE levels of 14 kJ.mol$^{-1}$, enabling rapid reactivity assessment without costly *ab initio* calculations. However, these models require substantial computational resources, which are unaffordable for many experimental labs, but also extensive data, which is challenging to acquire in chemistry considering data scarcity, heterogeneity, and cost. Nonetheless, effective simple models can be developed in low-data regime by constraining the chemical space. We have successfully developed such a high-performing oracle, achieving a MAE of less than 6 kJ.mol$^{-1}$ (i.e., less than 2% error on average). Moreover, while simple classical ML models may be less versatile than DL models, they can be adapted to extrapolate across different chemical spaces through careful feature selection. DL models continue to serve as powerful tools for predicting quantities with profoundly nonlinear relationships with the features. However, in the case of FIA, we have demonstrated that linear models based on physically meaningful descriptors perform exceptionally well, suggesting a linear relationship between FIA and molecular parameters.

Apart from their simplicity, linear models and decision trees, that use chemical features in a way that resembles chemical thought, are also more interpretable than “black-box” DL models. Models that can learn and represent genuine chemical concepts are essential for integration into a broader scientific approach: building models, making predictions, interpreting results for deeper understanding, and eventually refining the models. Developing such ML approaches grounded in physical reality remains a challenge in chemistry.[21]

In this study, we leveraged explainable ML in two key ways. We employed quantum descriptors based on electronic structure parameters and revealed that the reactivity of the studied boron Lewis acids is governed by molecular orbital interactions, classifying them as soft Lewis acids. However, these quantum descriptors are not readily actionable for molecular design. To address this, we introduced “Hammett-extended” descriptors, which focus on the nature and positioning of substituents. Interpretations from these models align with the XAI attributes identified by Wellawatte et al.[25] They are *actionable* and *succinct*, and, as they align with the language and concepts of organic chemistry, they are *domain applicable*. These explanations are also *correct*, corresponding to the expected electronic demand of a substituent at a given position. By interpreting models based on these descriptors, we unraveled the interdependent effects of substituents and identified design rules for creating novel ONO compounds with targeted Lewis acidity.

Here, our aim was to provide a route for the design of novel organic compounds, bridging the gap between pure ML techniques and traditional intuition-based strategies in organic chemistry. While our focus was centered on Lewis acidity, this methodology holds promise for exploring other types of reactivity, provided relevant parameters can be readily accessed (either computationally or experimentally) to construct a relevant dataset.

# Associated content

Supporting information is available for additional details and extended results (Lewis acidity metrics benchmark, computational methods, database construction, molecular descriptors implementation, machine learning models building and interpretation). This project was implemented in Python leveraging the Scikit-Learn and RDKit libraries. The code, including optimized models and datasets, is publicly available at https://github.com/jfenogli/XAI_boron_LA. This repository also includes Jupyter notebooks to reproduce the analyses presented in this manuscript and adapt workflows for new datasets or reactivity studies.

## Acknowledgement

The authors would like to thank Dr. Jules Schleinitz for his valuable guidance at the beginning of this project and his feedback on the manuscript. We also acknowledge Pablo Mas for his assistance with database extension via clustering algorithms and Maxime R. Vitale for fruitful discussion.

# Supporting information

## Contents

# S1 Lewis acidity metrics benchmark

As explained in the manuscript, we employed a computed quantity to account for the Lewis acidity of boron derivatives, the fluoride ion affinity (FIA). However, this choice results from a preliminary study on the different possible scales to account for Lewis acidity. This study is detailed here.

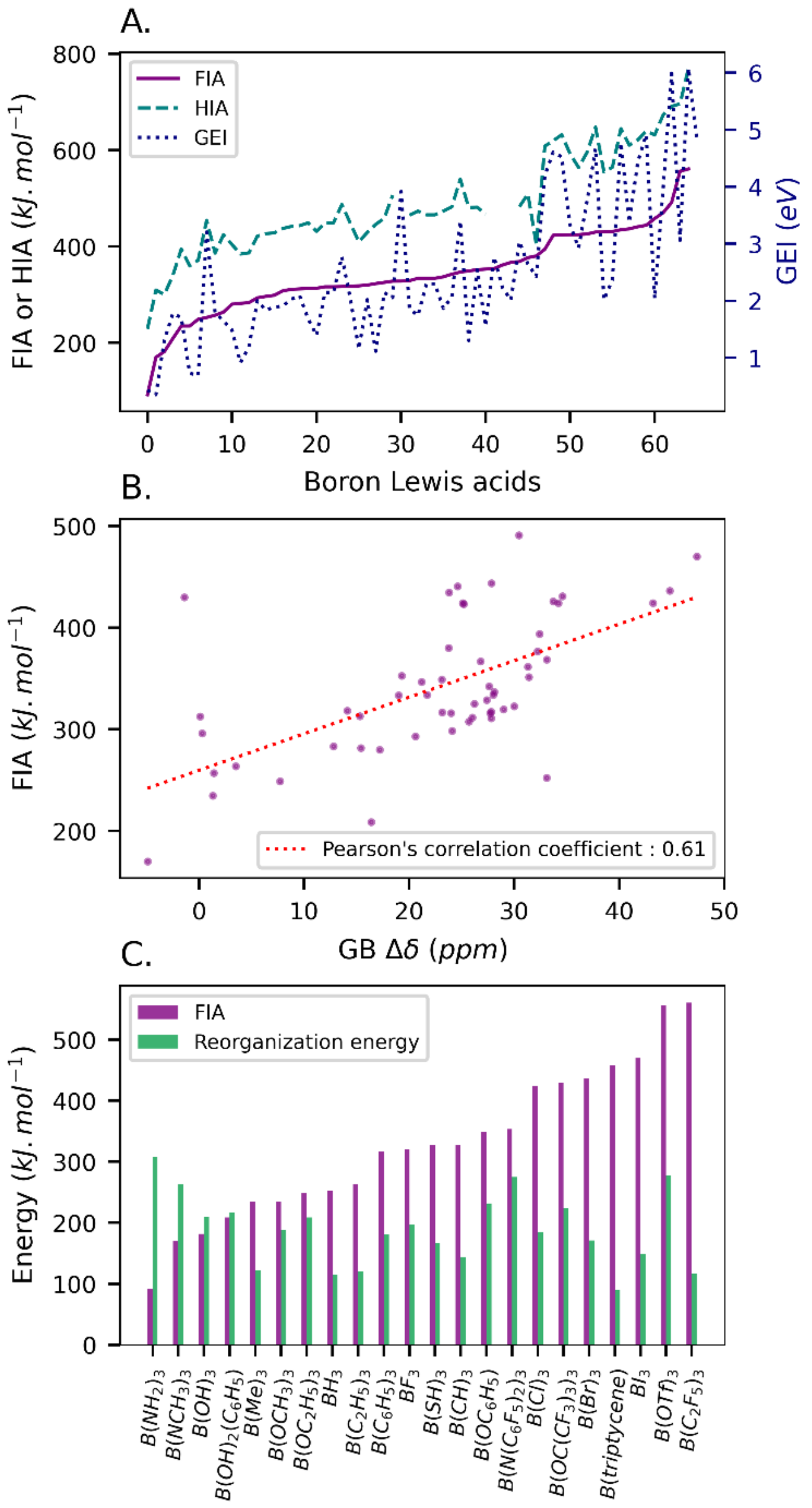

Figure S1 : Lewis acidity description: A. correlation between FIA, HIA and GEI, B. correlation between FIA and GB Δδ($^{31}$P), C. comparison of FIA and the corresponding reorganization energy for various common Lewis acids.

Traditional Lewis acidity scales typically involve measuring the interaction of a Lewis acid with a specific base for comparison. Among these scales, FIA is particularly well-established, as the fluoride ion offers significant advantages as a Lewis base. Its small size and low polarizability minimize steric hindrance and mitigate interfering effects such as charge transfer, π-back-bonding, and dispersion forces.[1] Experimental determination of FIA requires precise conditions[2] ; hence, it is frequently calculated computationally using density functional theory (DFT), ensuring consistency.

According to the Hard and Soft Acids and Bases (HSAB) theory,[3] the fluoride ion is classified as a hard base. Consequently, we also considered the hydride ion affinity (HIA) as an alternative Lewis acidity scale, given that the hydride ion is a soft Lewis base, although it is less commonly employed.

Both FIA and HIA are thermodynamic measures of Lewis acidity, based on the standard enthalpy change of Lewis adduct formation. In addition to these thermodynamic Lewis acidity scales,[4] it is also possible to assess a compound's Lewis acidity through analysis of its electronic structure, known as intrinsic Lewis acidity scales,[4] which require only minimal quantum calculation. One example is the global electrophilicity index (GEI),[5] that can easily be calculated from the energies of the molecule's frontier orbitals :

$$GEI = \frac{\chi^2}{2\eta}$$

where χ represents the molecule's global electronegativity and η denotes its hardness. The GEI measures a molecule's ability to accept electrons and has the advantage of being independent of the Lewis base. However, it is linked to reaction kinetics, contrasting with the conventional thermodynamic definition of Lewis acidity. Refer to part *S2 Computational methods* for details.

These theoretical scales should correlate with the experimental ones, even though experimental data can be noisy. Typical experimental methodologies involve measuring changes in the physicochemical properties of a probe base molecule upon binding with the studied Lewis acid. Two preferred bases are triethylphosphine oxide -or another phosphine oxide-, using the Gutmann-Beckett (GB) method (Figure S2.A),[4,6] and crotonaldehyde,[7] following Child's method (Figure S2.B). For both methods, the approach involves comparing nuclear magnetic resonance (NMR) chemical shifts, either of the terminal proton in crotonaldehyde or of the phosphorus atom in triethylphosphine oxide, when free in solution and when bound to the Lewis acid. The difference in NMR shifts, denoted as Δδ, serves as a consistent metric for comparing different Lewis acids.[4]

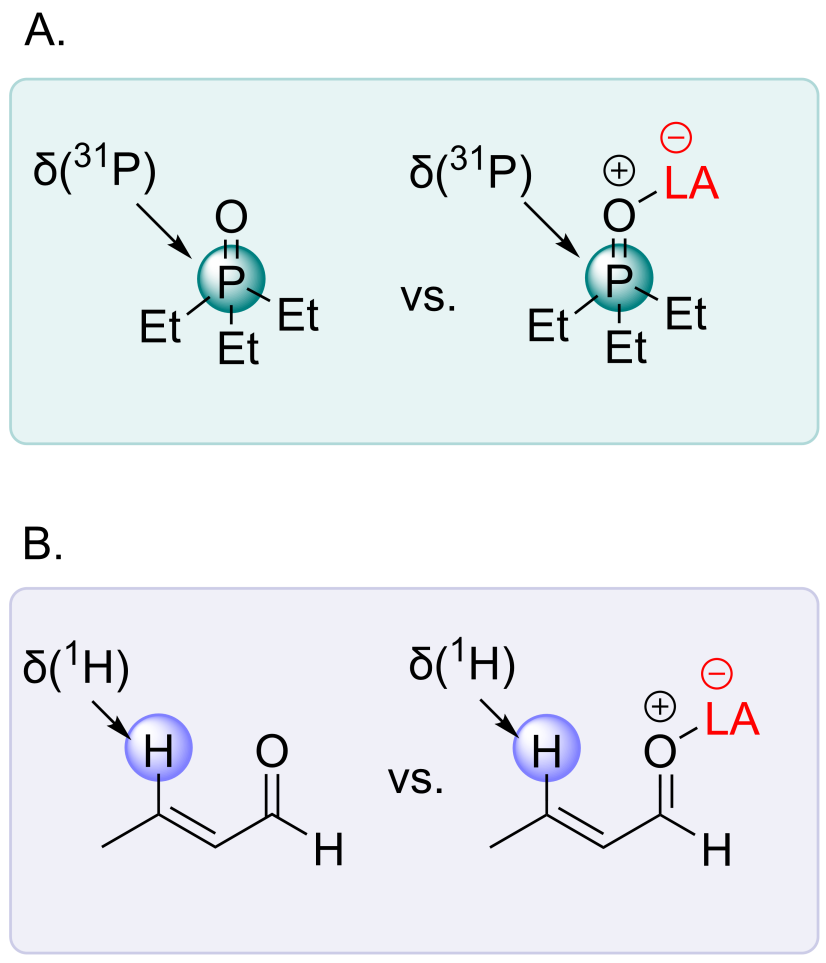

Figure S2 : NMR spectroscopy methodologies to determine the Lewis acidity of a Lewis acid (LA): A. Gutmann-Beckett method. B. Childs method.

Comparing Lewis acidity scales is complex, particularly as experimental measurements of Lewis acidity are prone to numerous artifacts. Challenges such as poor solubility or incomplete adduct formation can interfere with the accurate measurement of phosphorus chemical shift.[4] Moreover, considering the need for a large dataset of boron Lewis acids to build a machine learning (ML) database, having a computable quantity that can be consistently accessed across molecules is advantageous. To ensure alignment with the experimental scale, we first calculated the fluoride ion affinity (FIA), hydride ion affinity (HIA), and global electrophilicity index (GEI) for various Lewis acids (56 molecules) where the GB $\Delta\delta(^{31}P)$ was reported in literature (refer to Figure S1.B). The computed values and NMR data collected from literature are presented in Table S1 and analyzed in a pair-plot (Figure S3).

FIA or HIA and GEI are somewhat correlated but not perfectly (Figure S1.A); Pearson's correlation coefficient (Pearson's r) between FIA and GEI is 0.76, indicating that electrophilicity and Lewis acidity follow similar trends but are distinct concepts. FIA and HIA exhibit a strong correlation (Pearson's r = 0.95), allowing to use them interchangeably for comparing boron Lewis acids. Since the FIA is a global Lewis acidity scale and shows a reasonably strong correlation (Pearson's r = 0.61) with the Gutmann-Beckett $\Delta\delta(^{31}P)$ as shown in Figure S1.B, we selected it as the metric for labeling the Lewis acids of our database. The discrepancies between the FIA and GB scales primarily stem from steric effects, which are not fully captured by FIA. For instance, tris(pentachlorophenyl)borane ($B(C_6Cl_5)_3$) is predicted to be a strong Lewis acid with a computed FIA of 429.7 kJ.mol-1 ; however, its effective Lewis acidity measured by the GB method is only -1.4 ppm, attributed to steric hindrance.[8]

Berionni et al. have demonstrated that altering the typical planar trigonal boron coordination environment can significantly enhance the Lewis acidity of boranetriptycenes.[9] We have thus calculated the reorganization energy of Lewis acids, which quantifies the energy difference between the Lewis acid's geometry upon adduct formation (removing the fluoride ion) and its initial equilibrium geometry (Figure S1.C). As expected, Berionni's borane triptycene[9] achieves the smaller reorganization energy, due to the preorganized geometry of the ligand.

The lower the reorganization energy, the higher the expected FIA, which is supported by the negative correlation coefficient between reorganization energy and FIA. However, the correlation is weak (Pearson's r = -0.27, Figure S1.C). While reorganization energy significantly influences the Lewis acidity of a compound,[9] it is heavily dependent on the molecular structure. For a given ligand scaffold with various substitutions, the values fluctuate only slightly around the mean (Figure S4). Since our interest lies primarily in varying the substituents rather than the carbon backbone of the ligand, reorganization energy alone was not a suitable metric for our purposes. However, FIA, being the opposite of the reaction enthalpy of adduct formation, also incorporates the geometric reorganization energy associated with complexation, making it a more comprehensive and relevant parameter for assessing Lewis acidity.

| **SMILES** | **$\delta(^{31}P)$ (solvent)** | **$\Delta\delta(^{31}P)$** | **average $\Delta\delta(^{31}P)$** | **FIA (kJ.mol-1)** | **HIA (kJ.mol-1)** | **Reorganization energy (kJ.mol-1)** | **GEI (eV)** |
|---|---|---|---|---|---|---|---|
| **CCOB(OCC)OCC** | 48.7 (neat)[6] | 7.7[6] | 7.7 | 248.69 | 371.27 | 208.95 | 0.69 |
| **COB(OC)OC** | 48.1 (C6D6)[10] | 1.3[10] | 1.3 | 234.63 | 359.50 | 188.53 | 0.73 |
| **ClCCOB(OCCCl)OCCCl** | 55.1 (neat)[11] | 14.1[11] | 14.1 | 318.12 | 409.35 | 214.23 | 1.15 |
| **ClCCCOB(OCCCCl)OCCCCl** | 56.3 (neat)[11] | 15.3[11] | 15.3 | 312.98 | 431.19 | 221.68 | 1.39 |
| ClCCCCOB(OCCCCCl)OCCCCCl | 53.8 (neat)[11] | 12.8[11] | 12.8 | 283.23 | 386.44 | 237.24 | 1.19 |
| ClC(Cl)COB(OCC(Cl)Cl)OCC(Cl)Cl | 67.8 (neat)[11] | 26.8[11] | 26.8 | 366.71 | | 223.02 | 2.01 |
| ClC(Cl)(Cl)COB(OCC(Cl)(Cl)Cl)OCC(Cl)(Cl)Cl | 73.2 (neat)[11] | 32.2[11] | 32.2 | 376.53 | 508.65 | 247.59 | 2.64 |
| ClCC(CCl)OB(OC(CCl)CCl)OC(CCl)CCl | 60.3 (neat)[11] | 19.3[11] | 19.3 | 352.71 | 466.71 | 217.16 | 1.55 |
| BrCCOB(OCCBr)OCCBr | 58.2 (neat)[11] | 17.2[11] | 17.2 | 279.96 | 403.56 | 248.73 | 1.49 |
| ICCOB(OCCI)OCCI | 61.6 (neat)[11] | 20.6[11] | 20.6 | 293.11 | 421.32 | | 2.00 |
| FC(F)(F)COB(OCC(F)(F)F)OCC(F)(F)F | 71 (neat)[11] | 30[11] | 30 | 322.69 | 445.66 | | 1.11 |
| CC[Si](CC)(CC)OB(O[Si](CC)(CC)CC)O[Si](CC)(CC)CC | 56.4 (neat)[12] | 15.4[12] | 15.4 | 281.58 | 384.07 | 235.02 | 0.91 |
| Fc1c(F)c(F)c(B(c2c(F)c(F)c(F)c(F)c2F)c2c(F)c(F)c(F)c(F)c2F)c(F)c1F | 77 (CD2Cl2)[13], 75.4 (C6D6)[13] | 26.3[13], 29.4[13] | 27.85 | 443.82 | 639.34 | | 4.91 |
| c1ccc(B(c2ccccc2)c2ccccc2)cc1 | 72.5 (C6D6)[10] | 20.6[4], 25.7[10] | 23.15 | 316.68 | 486.91 | 181.31 | 2.77 |
| Fc1cc(F)c(F)c(B(c2c(F)c(F)cc(F)c2F)c2c(F)c(F)cc(F)c2F)c1F | 77.3 (CD2Cl2)[14] | 25.2[14] | 25.2 | 422.97 | 619.13 | | 4.61 |
| Fc1cc(B(c2cc(F)c(F)c(F)c2F)c2cc(F)c(F)c(F)c2F)c(F)c(F)c1F | 76.7 (CD2Cl2)[15] | 24.6[15] | 24.6 | 440.66 | 620.91 | | 4.44 |
| Fc1cccc(F)c1B(c1c(F)cccc1F)c1c(F)cccc1F | 72.6 (CD2Cl2)[15] | 21.2[15] | 21.2 | 346.60 | 539.54 | | 3.40 |
| FC(F)(F)c1cc(B(c2cc(C(F)(F)F)cc(C(F)(F)F)c2)c2cc(C(F)(F)F)cc(C(F)(F)F)c2)cc(C(F)(F)F)c1 | 78.9(CD2Cl2)[16] | 28.2[16] | 28.2 | | | | 4.82 |
| Fc1c(F)c(F)c(B(c2c(F)c(F)c(F)c(F)c2F)c2c(Cl)c(Cl)c(Cl)c(Cl)c2Cl)c(F)c1F | 75.8 (CD2Cl2)[8] | 25.1[8] | 25.1 | 424.08 | 631.72 | | 4.53 |
| Fc1c(F)c(F)c(B(c2c(Cl)c(Cl)c(Cl)c(Cl)c2Cl)c2c(Cl)c(Cl)c(Cl)c(Cl)c2Cl)c(F)c1F | 74.5 (CD2Cl2)[8] | 23.8[8] | 23.8 | 434.46 | 644.34 | | 4.77 |
| FC(F)(F)c1cc(C(F)(F)F)c(Bc2c(C(F)(F)F)cc(C(F)(F)F)cc2C(F)(F)F)c(C(F)(F)F)c1 | 78.7(C6D6)[17] | 32.4[17] | 32.4 | 393.92 | 607.95 | | 4.20 |
| c1ccc(OB(Oc2ccccc2)Oc2ccccc2)cc1 | 69.45(C6D6)[18] | 23.1[18] | 23.1 | 349.03 | 480.62 | 231.64 | 1.30 |
| Fc1c(F)c(F)c(OB(c2c(F)c(F)c(F)c(F)c2F)c2c(F)c(F)c(F)c(F)c2F)c(F)c1F | 80(C6D6)[18] | 33.7[18] | 33.7 | 425.67 | 595.22 | | 3.75 |
| Fc1c(F)c(F)c(OB(Oc2c(F)c(F)c(F)c(F)c2F)c2c(F)c(F)c(F)c(F)c2F)c(F)c1F | 80.5(C6D6)[18] | 34.2[18] | 34.2 | 424.26 | 563.69 | | 2.93 |
| Fc1c(F)c(F)c(OB(Oc2c(F)c(F)c(F)c(F)c2F)Oc2c(F)c(F)c(F)c(F)c2F)c(F)c1F | 80.9(C6D6)[18] | 34.6[18] | 34.6 | 430.73 | 564.14 | | 2.37 |
| FB(F)F | 80.9 (neat)[4] | 29[4] | 29 | 319.65 | 428.92 | 197.18 | 2.02 |
| FB(F)OOOSC(F)(F)F | 84.6 (CDCl3)[19] | 33.1[19] | 33.1 | 368.45 | 482.43 | | 3.04 |
| B | | 33.1[4] | 33.1 | 252.30 | 453.88 | 114.25 | 3.21 |
| CCB(CC)CC | 51.9 (C6D6)[10] | 1.9[4], 5.1[10] | 3.5 | 263.54 | 424.45 | 120.61 | 1.64 |
| ClB(Cl)Cl | 88.7 (neat)[6] | 47.7[6], 38.7[4] | 43.2 | 424.23 | 591.88 | 184.33 | 3.33 |
| BrB(Br)Br | 90.3 (neat)[6] | 49.3[6], 40.3[4] | 44.8 | 436.41 | 610.00 | 170.92 | 3.38 |
| IB(I)I | 92.9 (neat)[6] | 51.9[6], 42.9[4] | 47.4 | 469.81 | 672.95 | 149.09 | 3.97 |
| OB(O)c1ccccc1 | 63.2(C6D6)[10] | 16.4[10] | 16.4 | 208.73 | 339.53 | 216.78 | 1.78 |
| **c1ccc(B2Oc3ccccc3O2)cc1** | 70.5(C6D6)[20] | 24.1[20] | 24.1 | 298.45 | 427.99 | 188.20 | 1.89 |
| **Fc1ccccc1B1Oc2ccccc2O1** | 74.2(C6D6)[20] | 27.8[20] | 27.8 | 310.75 | 441.25 | | 2.03 |
| **Fc1cccc(B2Oc3ccccc3O2)c1** | 72.4(C6D6)[20] | 26[20] | 26 | 311.66 | 442.95 | | 2.09 |
| **Fc1ccc(B2Oc3ccccc3O2)cc1** | 72.1(C6D6)[20] | 25.7[20] | 25.7 | 307.55 | 437.03 | | 1.90 |
| **Fc1ccc(B2Oc3ccccc3O2)c(F)c1** | 74.2(C6D6)[20] | 27.8[20] | 27.8 | 317.25 | 447.72 | | 2.04 |
| **Fc1cccc(F)c1B1Oc2ccccc2O1** | 74.1 (C6D6)[20] | 27.7[20] | 27.7 | 316.04 | 448.68 | | 2.15 |
| **Fc1cc(B2Oc3ccccc3O2)cc(F)c1F** | 74.4 (C6D6)[20] | 28[20] | 28 | 333.67 | 465.44 | | 2.30 |
| **Fc1cc(F)c(B2Oc3ccccc3O2)c(F)c1** | 72.6 (C6D6)[20] | 26.2[20] | 26.2 | 325.11 | 457.94 | | 2.14 |
| **Fc1c(F)c(F)c(B2Oc3ccccc3O2)c(F)c1F** | 77.8 (C6D6)[20] | 31.4[20] | 31.4 | 351.40 | 481.32 | | 2.54 |
| **Fc1c(F)c(F)c(B2Cc3ccc4ccccc4c3-c3c<br>(ccc4ccccc34)C2)c(F)c1F** | 72.7 (CD2Cl2)[13],<br>71.2 (C6D6)[13] | 22[13],<br>25.5[13] | 23.75 | 380.03 | 403.61 | | 2.43 |
| **CC1(C)OB(c2ccccc2F)OC1(C)C** | 47.8 (C6D6)[20] | 1.4[20] | 1.4 | 257.02 | 385.94 | | 1.79 |
| **Fc1c(F)c(C(F)(F)F)c(F)c(F)c1B(c1c(F)c(F)<br>c(C(F)(F)F)c(F)c1F)c1c(F)c(F)c(C(F)(F)F)c(F)c1F** | 79.5 (CD2Cl2)[21],<br>77.7 (C6D6)[21] | 29[21],<br>31.9[21] | 30.45 | 490.99 | 690.76 | | 5.99 |
| **Clc1c(Cl)c(Cl)c(B(c2c(Cl)c(Cl)c(Cl)c(Cl)<br>c2Cl)c2c(Cl)c(Cl)c(Cl)c(Cl)c2Cl)c(Cl)c1Cl** | 50.7 (CD2Cl2)[8] | -1.4[8] | -1.4 | 429.70 | 647.31 | 157.18 | 4.67 |
| **CN(C)B(N(C)C)N(C)C** | | -4.9[4] | -4.9 | 169.83 | 309.15 | 262.83 | 0.35 |
| **Fc1cc(B2Oc3ccccc3O2)cc(F)c1F** | 67.9(C6D6)[22] | 21.7[22] | 21.7 | 333.67 | 465.44 | | 2.30 |
| **Fc1c(F)c(F)c(B2Oc3cccc4cccc(c34)O2)c(F)c1F** | 77.5(C6D6)[22] | 31.3[22] | 31.3 | 361.31 | 495.77 | | 2.22 |
| **Fc1cc(B2Oc3cccc4cccc(c34)O2)cc(F)c1F** | 73.8(C6D6)[22] | 27.6[22] | 27.6 | 342.18 | 481.96 | | 2.06 |
| **Fc1cc(F)c(B2Oc3cccc4cccc(c34)O2)c(F)c1** | 74.3 (C6D6)[22] | 28.1[22] | 28.1 | 336.77 | 473.19 | | 1.86 |
| **Fc1cccc(F)c1B1Oc2cccc3cccc(c23)O1** | 73.6 (C6D6)[22] | 27.4[22] | 27.4 | 328.77 | 464.73 | | 1.85 |
| **Clc1cccc(Cl)c1B1Oc2cccc3cccc(c23)O1** | 65.2 (C6D6)[22] | 19[22] | 19 | 333.59 | 473.11 | 187.88 | 1.74 |
| **Cc1cc(C)c(B2Oc3cccc4cccc(c34)O2)c(C)c1** | 46.3 (C6D6) [22] | 0.1[22] | 0.1 | 312.71 | 448.76 | 176.74 | 1.68 |
| **Fc1cccc(F)c1B1Oc2ccccc2O1** | 70.2 (C6D6)[22] | 24[22] | 24 | 316.04 | 448.68 | | 2.15 |
| **Cc1cc(C)c(B2Oc3ccccc3O2)c(C)c1** | 46.5 (C6D6)[22] | 0.3[22] | 0.3 | 296.11 | 426.28 | 163.71 | 1.83 |
| **C#CB(C#C)C#C** | | | | 327.86 | | 144.37 | 3.92 |
| **c1ccc2c(c1)B1c3ccccc3C2c2ccccc21** | | | | 457.77 | 631.25 | 89.73 | 2.04 |
| **CB(C)C** | | | | 234.60 | 394.51 | 122.14 | 1.71 |
| **FC(F)(F)C(F)(F)B(C(F)(F)C(F)(F)F)C(F)(F)C(F)(F)F** | | | | 560.68 | 769.73 | 117.24 | 6.08 |
| **FC(F)(F)C(OB(OC(C(F)(F)F)(C(F)(F)F)C(F)(F)F)O<br>C(C(F)(F)F)(C(F)(F)F)C(F)(F)F)(C(F)(F)F)C(F)(F)F** | | | | 429.88 | 554.52 | 224.35 | 2.01 |
| **Fc1c(F)c(F)c(N(B(N(c2c(F)c(F)c(F)c(F)c2F)c2<br>c(F)c(F)c(F)c(F)c2F)N(c2c(F)c(F)c(F)c(F)c2F)<br>c2c(F)c(F)c(F)c(F)c2F)c2c(F)c(F)c(F)c(F)c2F)c(F)c1F** | | | | 353.84 | | 275.20 | 2.77 |
| **NB(N)N** | | | | 92.01 | 228.71 | 307.55 | 0.43 |
| **O=S(=O)(OB(OS(=O)(=O)C(F)(F)F)OS(=O)(=O)C(F)<br>(F)F)C(F)(F)F** | | | | 555.93 | 695.88 | 276.97 | 3.06 |
| **OB(O)O** | | | | 181.47 | 300.20 | 209.07 | 1.31 |
| **SB(S)S** | | | | 327.73 | 506.10 | 166.85 | 2.05 |

Table S1 : Lewis acidity data for a bunch of Lewis acids. GB $\Delta\delta(^{31}P)$ calculated from NMR spectroscopy data reported in the literature, computed FIA, reorganization energy and GEI at the DFT M062X/6-31Gd level of theory. As $\delta(^{31}P)$ of the triethylphosphine oxide complexed with the Lewis acid were often reported in diverse solvents (mainly $CD_2Cl_2$ and $C_6D_6$)

and sometimes in neat conditions, we calculated the Δδ($^{31}$P) subtracting the value of δ($^{31}$P) for the triethylphosphine oxide in the appropriate conditions (as described in the GB methodology)[6]: 41 ppm (neat), 50.7 ppm ($CD_2Cl_2$), 46.3 ppm ($C_6D_6$) or 51.5 ppm ($CDCl_3$). When Δδ($^{31}$P) was also reported, this value was used instead. Then, the average value of all the Δδ($^{31}$P) reported for a same molecule was calculated and used for the following correlations.

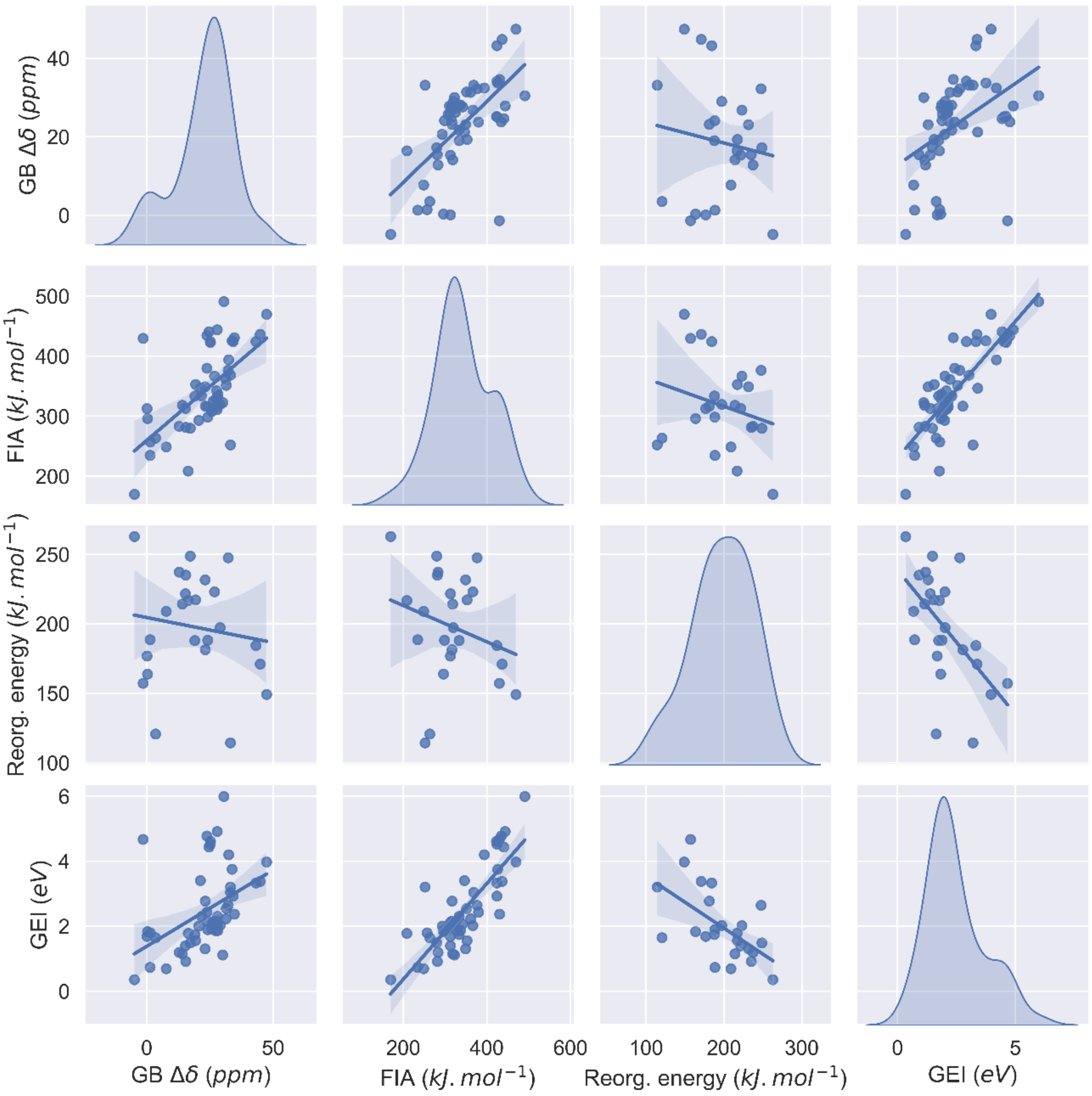

Figure S3 : pair-plot of different metrics used to quantify the Lewis acidity

| | average Δδ | FIA | HIA | reorg. energy | GEI |
|---|---|---|---|---|---|
| **average Δδ** | 1.00 | 0.61 | 0.54 | -0.14 | 0.44 |
| **FIA** | 0.61 | 1.00 | 0.95 | -0.27 | 0.76 |
| **HIA** | 0.54 | 0.95 | 1.00 | -0.43 | 0.87 |
| **reorg. energy** | -0.14 | -0.27 | -0.43 | 1.00 | -0.47 |

| GEI | 0.44 | 0.76 | 0.87 | -0.47 | 1.00 |
|---|---|---|---|---|---|

Table S2 : Correlation matrix (Pearson's r)

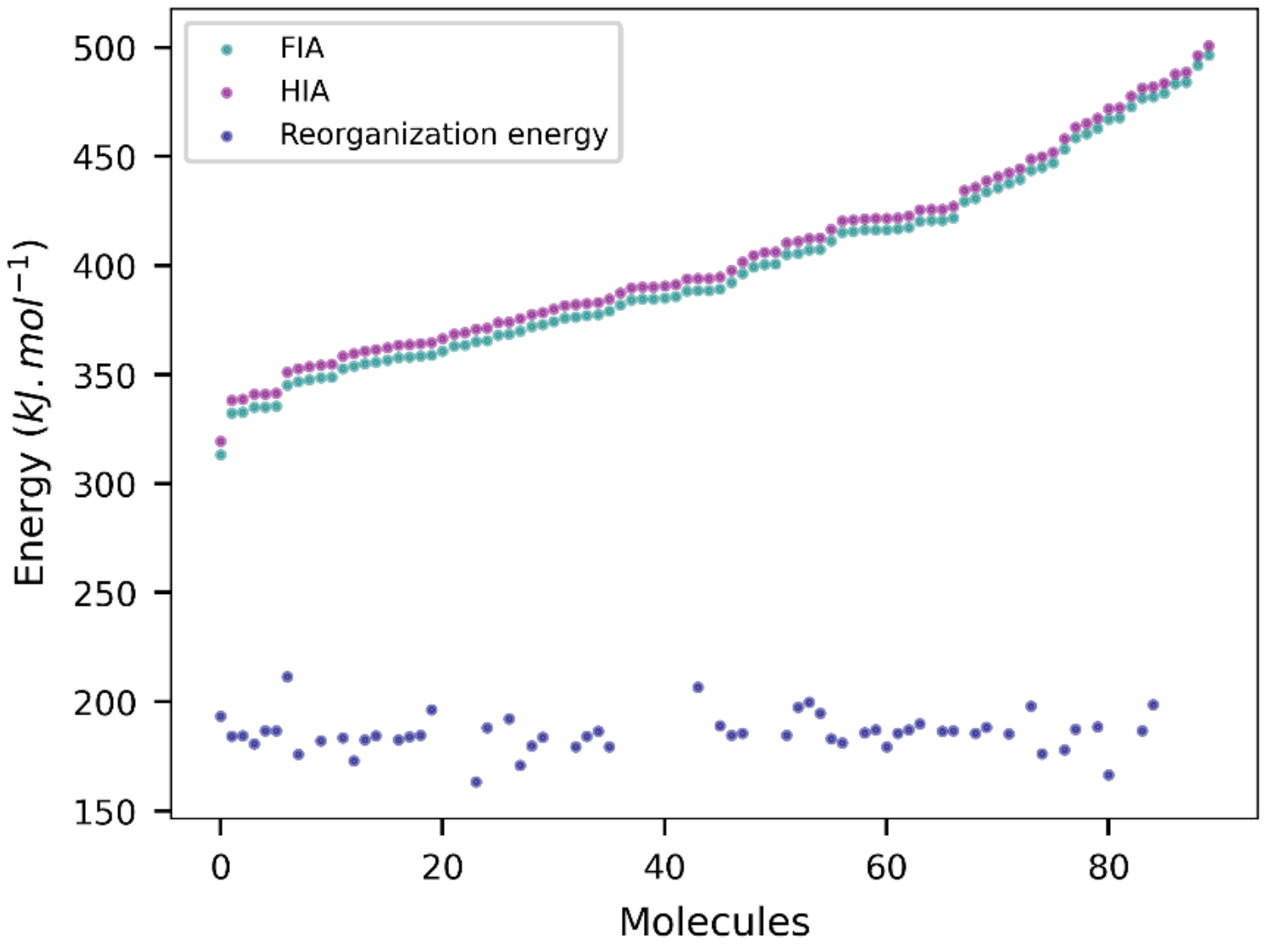

Figure S4 : FIA, HIA and reorganization energy for the ONO dataset; reorganization energy is almost constant across various functional groups substitutions on the scaffold (mean = 186 $kJ.mol^{-1}$, standard deviation = 8 $kJ.mol^{-1}$).

# S2 Computational methods

## FIA computation

### *Gas phase versus solution phase calculations*

Greb and colleagues have shown a strong correlation (Pearson's squared coefficient $r^2$ = 0.921) between FIA values calculated in the gas phase and those calculated in solution phase using dichloromethane as a solvent, based on their extensive dataset of 44k molecules.[23] This correlation is even stronger ($r^2$ = 0.941) for the smaller dataset of trivalent boron Lewis acids. Given that our reduced dataset also consists of boron Lewis acids with similar scaffolds, we expect an even higher correlation, allowing for a relevant comparison of FIA values in the gas phase.

### *Isodesmic calculations*

To efficiently constitute a database of several compounds, we must employ a relatively low-cost and yet precise *ab initio* method. DFT is the typical technique employed to compute efficiently molecular properties at the *ab initio* level. However, standard DFT methods do not accurately account for electronic correlation, leading to discrepancies when compared to non-approximated *ab initio* techniques. Usually, electronic correlation errors are canceled when calculating reaction enthalpies of similar molecules. However, in case of FIA, the involvement of a "naked" fluoride ion leads to differences in the electronic correlation with respect to the molecular Lewis acid or the Lewis acid–base adduct, that are not canceled out when computing the reaction enthalpy. This is why it is recommended to employ a fluoride ion donor such as $COF_3^-$ or $Me_3SiF$ as a reference system and proceed in two steps: determining the reaction enthalpy between the reference system and the Lewis acid studied in the gas phase (Scheme 1, Eq. (a)), and then subtracting the dissociation enthalpy of the reference system (Scheme 1, Eq. (b)).[24,25] This approach, known as “isodesmic calculations”, has been efficiently applied to reproduce FIA computed at the coupled-cluster level of theory by the Greb group.[1] The dissociation enthalpy of the reference system must be very precise, obtained either experimentally or with a high-level *ab initio* calculation. We used the value provided by Greb et al., 952.5 kJ.mol$^{-1}$, obtained through CCSD(T)/CBS theory for our calculations.[1] On the other hand, the calculation of the reaction enthalpy (a) can be performed at a more flexible calculation level, as long as the precision is sufficient.

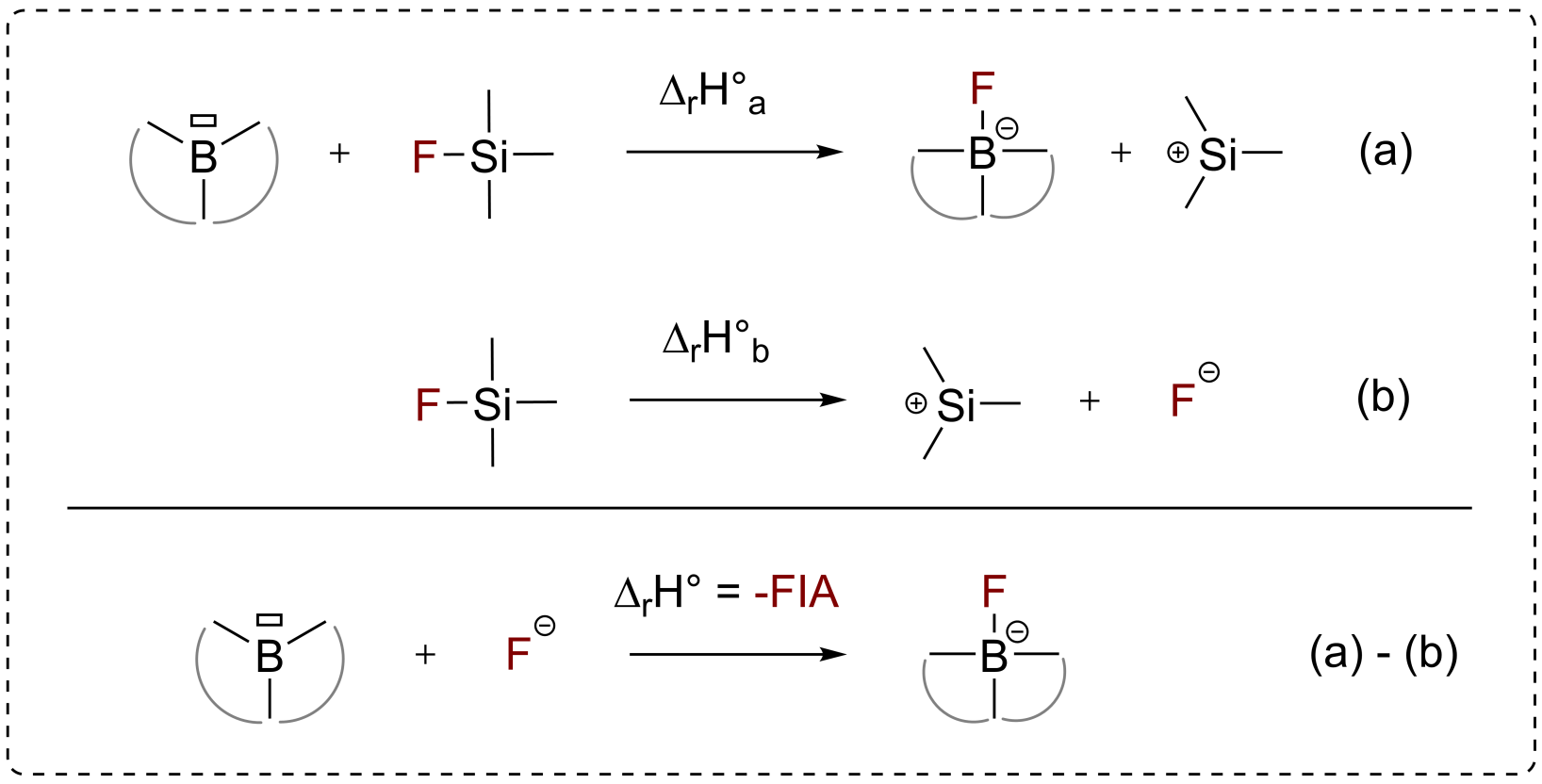

Scheme 1: Isodesmic reactions used for FIA calculation (all compounds are in the gas phase).

### *Choice of the DFT method*

Various *ab initio* methods, mainly isodesmic, were tested on a varied set of boron-based Lewis acids, and the obtained values were compared to those tabulated by Greb et al.,[1] which were obtained at a higher level of calculation (Figure S5 and Figure S6). The DFT methods proposed by Greb et al., particularly PBEP86, converge too slowly, especially when the chosen basis set is large (aug-cc-pVTZ). Therefore, we turned to less costly methods with a MAE of the same order of magnitude (Table S1), such as M052X and M062X, along with the 6-31G(d) basis set, which had already been reported in the literature for FIA calculations.[26]

Regarding the linear correlations shown in Figure S6, ideally, the slope should be 1 (which is roughly verified) and the intercept should be 0. However, a non-zero intercept only introduces a systematic bias and still allows us to observe trends, which allows the relative comparison of Lewis acids with one another. Since the performance of the M062X method is slightly superior to that of the M052X method (refer to the MAE values in Table S3), we chose M062X to generate the FIA data needed for database construction.

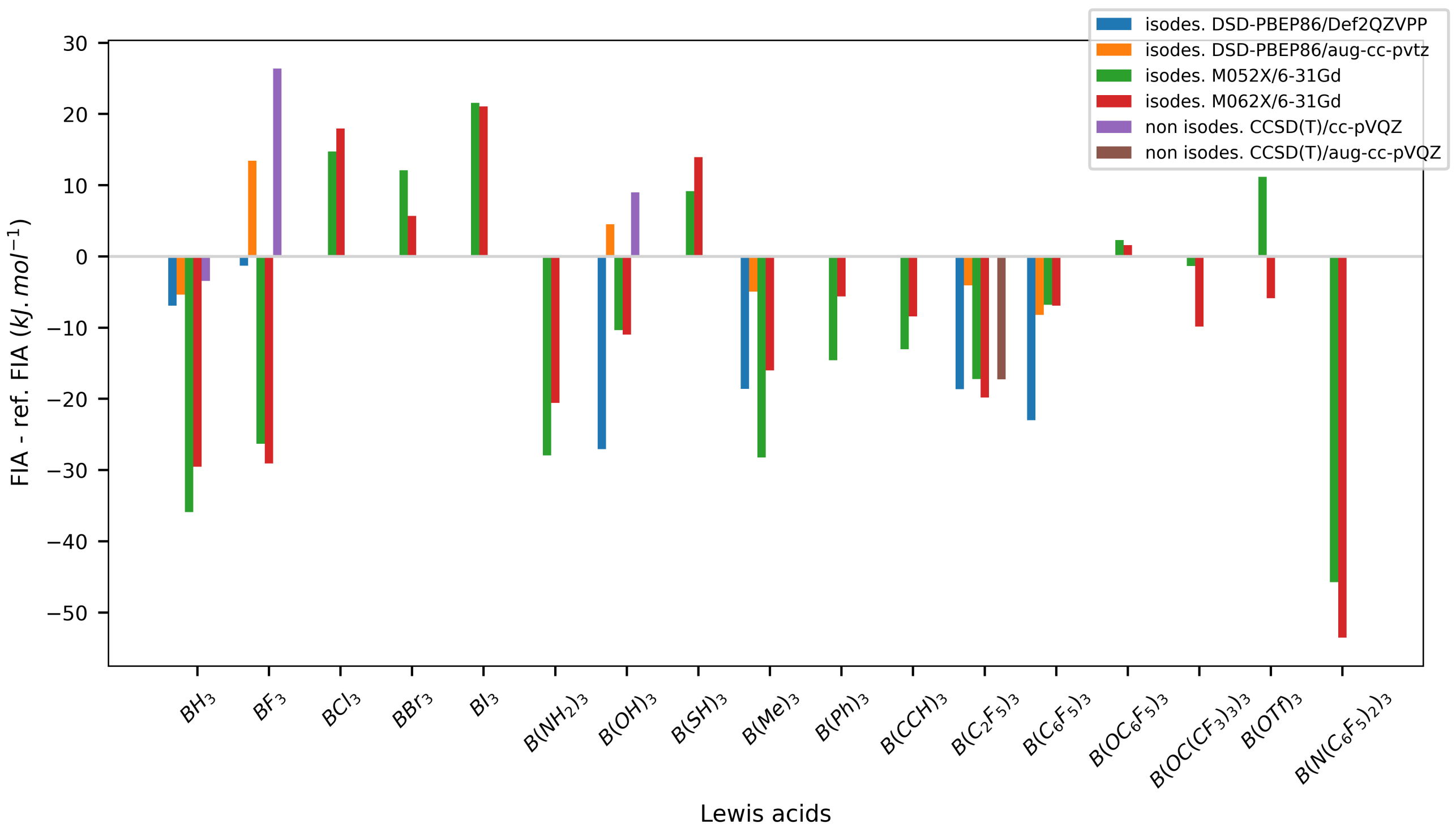

Figure S5 : FIA values relative to the reference values for each species and method.

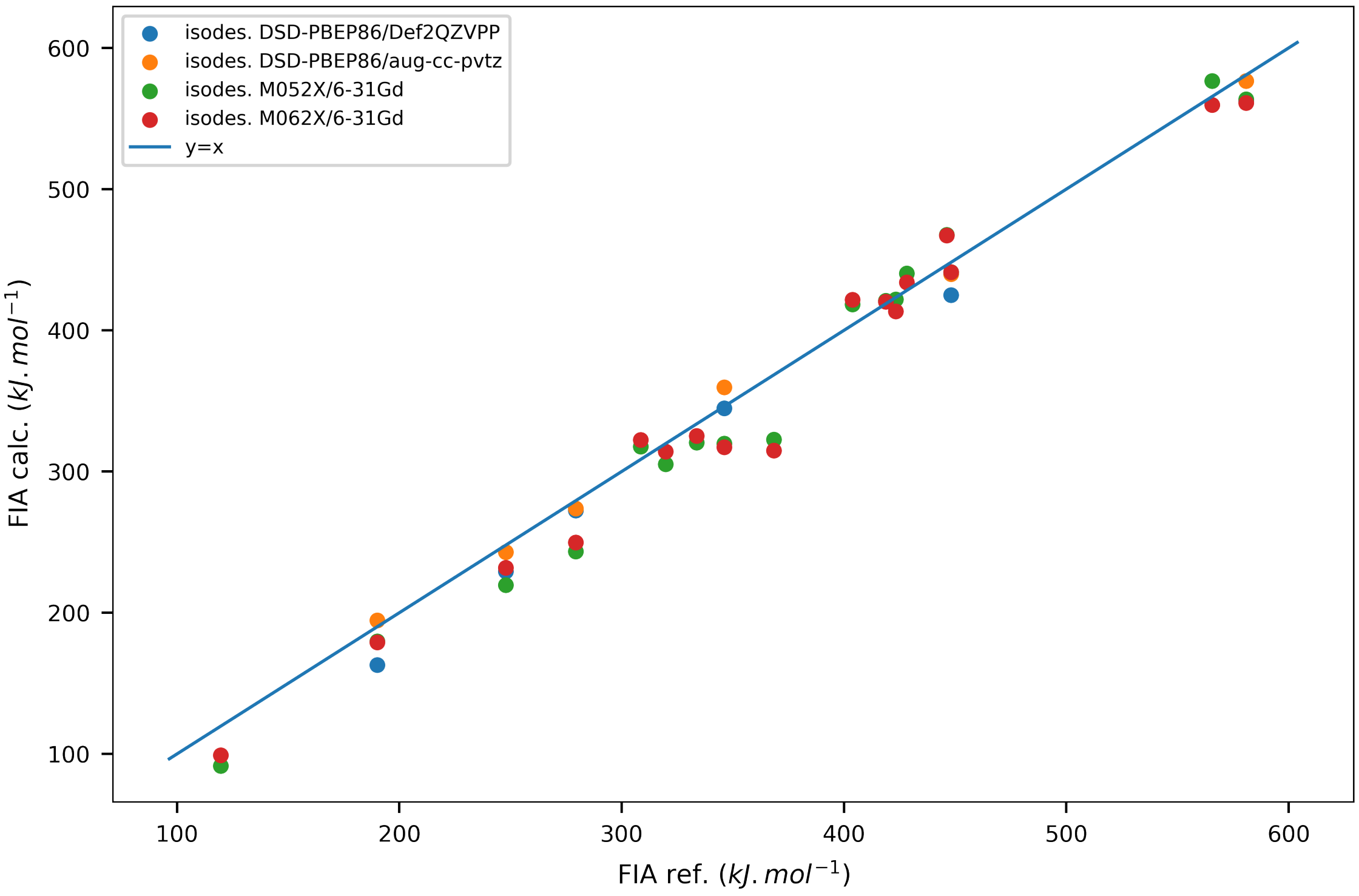

Figure S6 : Linear correlation between the calculated FIA values and the reference values.

| Method | isodes. DSD-PBEP86/def2QZVPP | isodes. DSD-PBEP86/aug-cc-pVTZ | isodes. M052X/6-31Gd | isodes. M062X/6-31Gd |
|---|---|---|---|---|
| **Slope** | 1 | 0.98 | 1.07 | 1.03 |
| **Intercept** | -16.07 | 4.74 | -36.62 | -21.77 |
| **$R^2$** | 0.995 | 0.997 | 0.982 | 0.979 |
| **MAE** | 7.89 | 5.78 | 13.42 | 13.1 |

Table S3 : parameters of the linear correlations plotted in Figure S6 and MAE

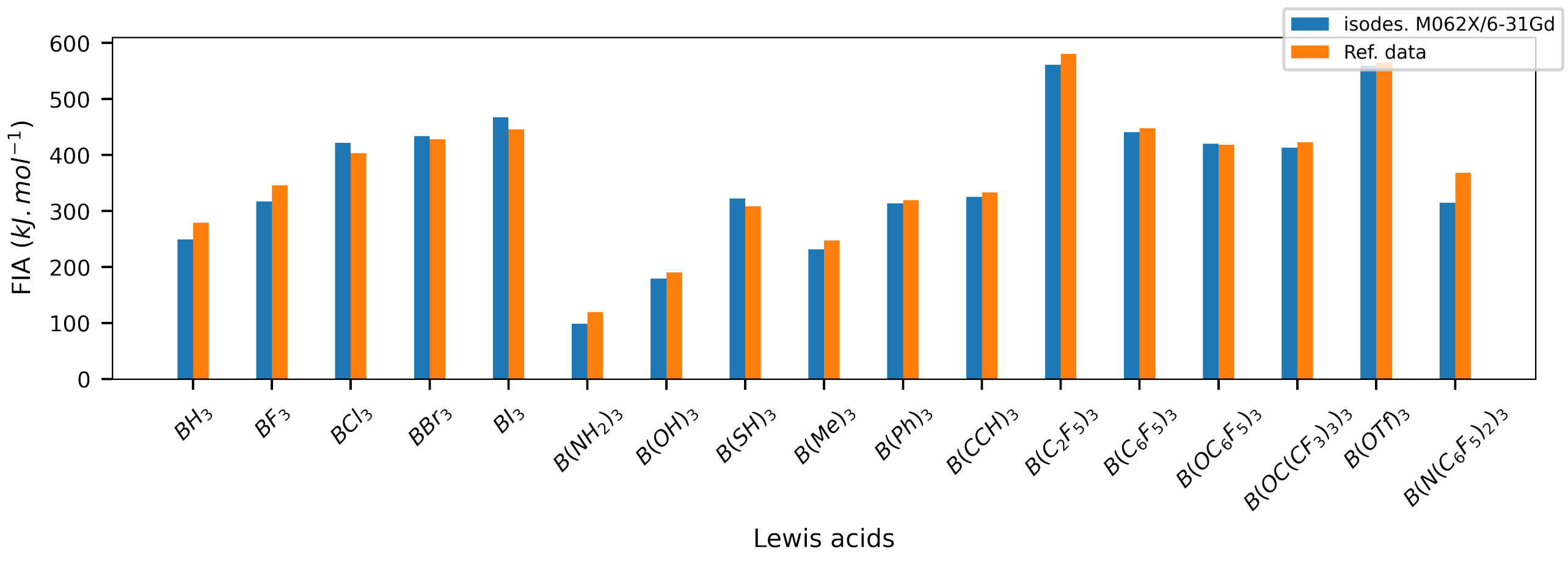

Figure S7 : FIA values calculated (M062X/6-31G(d)) and reference FIA values (from higher-level calculations) found in the literature for a set of boron-based Lewis acids.

## HIA computation

HIA data were computed primarily for comparison with FIA. For a given molecular scaffold, FIA and HIA are highly correlated (Figure S4). Unlike FIA, HIA was computed without using isodesmic calculations.

## GEI computation

The global electrophilicity index (GEI) was calculated from frontier molecular orbital energies ($E_{HOMO}$ and $E_{LUMO}$) obtained at the M062X/6-31G(d) level of theory with:

$$GEI = \frac{\chi^2}{2\eta}$$

Where, the global electronegativity can be calculated as $\chi = \left( \frac{E_{HOMO} + E_{LUMO}}{2} \right)$ and the hardness can be calculated as $\eta = \left( E_{LUMO} - E_{HOMO} \right)$.

## Reorganization energy computation

Reorganization energy calculations are conceptually trivial. The energy of reorganization is calculated in the following way: the fluoride ion of the borate is removed and a single point energy (SPE) calculation is performed on the boron derivative bended geometry obtained. The optimized derivative energy is subtracted to this quantity to yield the reorganization energy. Yet, as lots of SPE calculations failed (blanks in table S1), we computed less reorganization energies compared to the other metrics and did not pursue in computing missing values.

# S3 Chemical space

## Database extension

To the prediction performances of the models, we aimed to extend the database to at least 200 molecules. To achieve this, we performed a database extension for the ONO scaffold using the K-means[27] and coverage[28] clustering algorithms, targeting 50 clusters for each algorithm. These algorithms were applied to a Morgan fingerprint representation of 2,197 possible molecules in the chemical space, with the constraint of excluding molecules already present in the database. The K-means algorithm was repeated 10 times, while the coverage algorithm, that requires a longer convergence time, was repeated 4 times. From these runs, we selected the 50 molecules that appeared most frequently across all repetitions of both algorithms. Figure S8 shows the Tanimoto Similarity distribution of the selected molecules, which displays a relatively broad Gaussian distribution centered around a similarity of 0.35. This similarity value suggests a satisfying diversity in the selected molecular structures, indicating that these clustering algorithms effectively sampled the chemical space in a uniform manner. In total, we obtained 97 unique molecules (as 3 molecules were common to both algorithms), for which we calculated the FIA. As the K-means clustering algorithm was faster to process, we preferred it to extend the database of the triarylboranes of 100 molecules, for molecular design purposes.

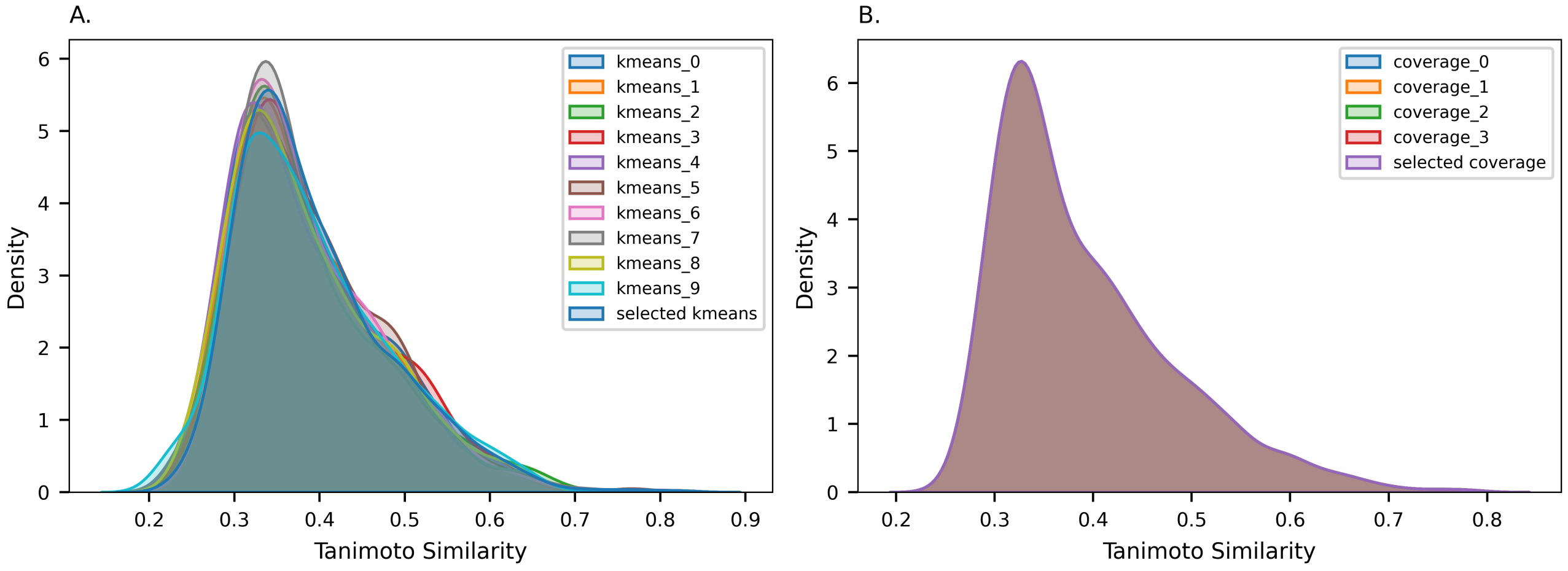

Figure S8 : Tanimoto similarity density curves of selected molecules for database extension for ONO scaffold, A. results for the K-means clustering algorithm, B. results for the coverage clustering algorithm.

## Database curation

FIA values were computed for each molecule and compiled into tables where molecules are represented as SMILES[29] strings. Outliers were detected plotting the distribution of FIA values and conducting PCA on quantum descriptors. These outliers were then eliminated from the database.

# S4 Molecular descriptors

## Chemo-informatics descriptors

We used the extended connectivity Morgan Fingerprints[30] and RDKit descriptors.[31,32] Fingerprints are 1024 bit-vectors emphasizing molecular connectivity and fragment diversity. RDKit descriptors provide structural and property-based features, including fragment counts, minimum and maximum absolute partial charges, molecular weight, and other QSAR descriptors (208 features in total).

## Quantum descriptors

Quantum features for each molecule were derived from the DFT calculation performed for FIA computation, using the Auto-QChem[33] workflow developed by the Doyle Group. This enabled efficient extraction of precise and physically meaningful features (amounting to 43), including global molecular properties (e.g., frontier orbitals energies) as well as local atomic properties for the boron atom only (e.g., partial charge). In spite of requiring more computational resources to compute, these descriptors offer reliable insights into the relationship between the quantum features of molecules and their Lewis acidity (see the manuscript, *Interpretability* section).

## Hammett-extended descriptors

These descriptors were generated by identifying the chemical nature of substituents at *ortho*, *meta* and *para* positions of each scaffold using the SMARTS substructure identifiers[34] implemented in the RDKit python library.[31] The code to generate the descriptors is available on the GitHub repository of the project.[35] Then, Hammett-extended parameters derived by Sigman and co-workers[36] corresponding to the *ortho*, *meta* and *para* substituents were concatenated into a vector featuring the molecule (36 features). These parameters, characteristic of the substituents on the benzoic acid, are the Sterimol parameters, $B1$, $B5$ and $L$ (in Å), infrared (IR) spectroscopy frequencies, $v$ (in $cm^{-1}$) and intensities $I$, Natural Bonding Orbital (NBO) charges (in e, elementary charge), and torsional angle between the aromatic ring plane and the carboxylic moiety, $\theta_{tor}$ (in °) (see Table S4, Table S5, and Table S6).

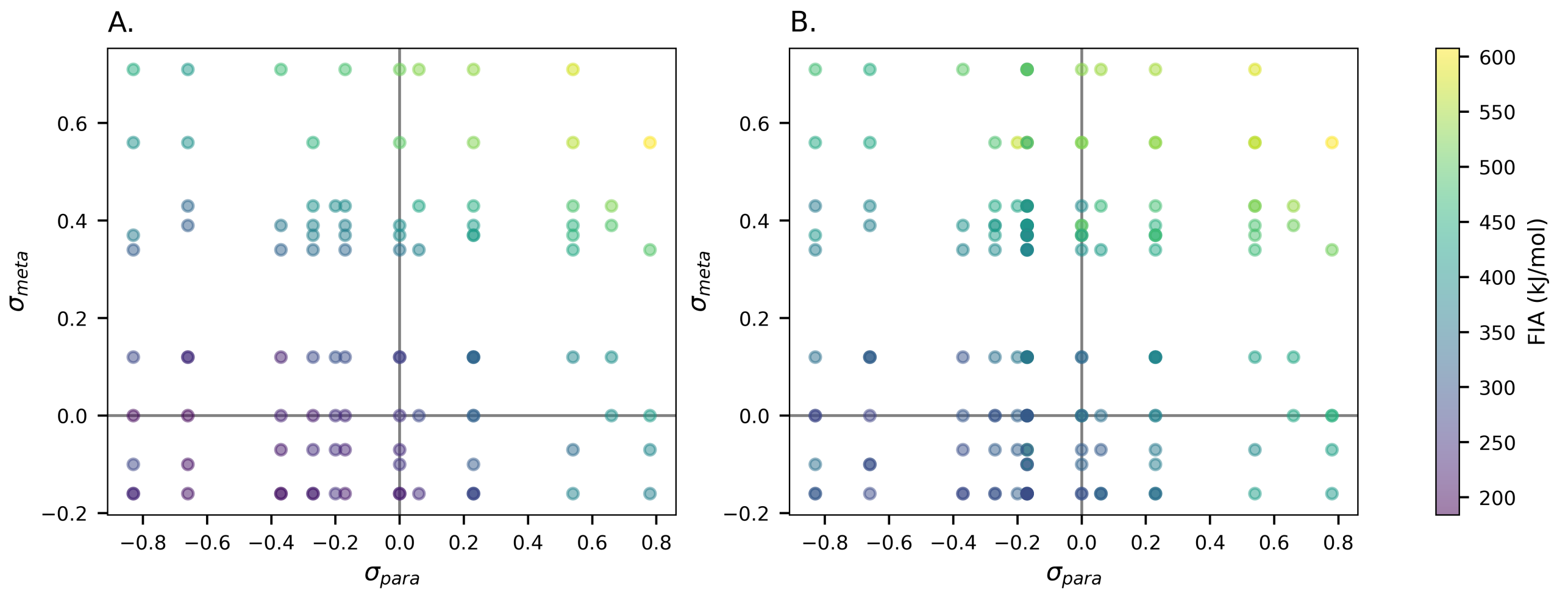

Figure S9 : Representations of the triarylboranes dataset following the value of their *para* and *meta* σ constants. A. Molecules from the triaryboranes dataset with no substituent in the *ortho* position. B. All molecules from the triarylborane dataset.

| *R-ortho* | $B_{1,o}$ | $B_{5,o}$ | $L_o$ | $\nu_{COH,o}$ | $I_{COH,o}$ | $\nu_{C=O,o}$ | $I_{C=O,o}$ | $NBO_{C,o}$ | $NBO_{=O,o}$ | $NBO_{O,o}$ | $NBO_{H,o}$ | $\theta_{tor}$ |
|---|---|---|---|---|---|---|---|---|---|---|---|---|
| **H** | 1.00 | 1.00 | 2.09 | 1394.60 | 163.25 | 1847.85 | 390.68 | 0.80516 | -0.60208 | -0.70358 | 0.50087 | 0.00 |
| **$NMe_2$** | 1.55 | 3.27 | 4.30 | 1385.34 | 46.99 | 1826.81 | 455.50 | 0.80697 | -0.61279 | -0.69946 | 0.49588 | 25.77 |
| **$NH_2$** | 1.55 | 2.06 | 3.03 | 1388.97 | 76.63 | 1830.53 | 481.33 | 0.80849 | -0.61276 | -0.73355 | 0.50600 | 2.60 |
| **OH** | 1.52 | 2.07 | 2.83 | 1408.63 | 122.13 | 1825.36 | 430.86 | 0.80852 | -0.61647 | -0.68045 | 0.49706 | 0.00 |
| **OCH3** | 1.52 | 3.22 | 4.25 | 1399.72 | 131.89 | 1837.11 | 429.50 | 0.81217 | -0.60468 | -0.69401 | 0.49943 | 21.20 |
| ***t*Bu** | 2.91 | 3.35 | 4.34 | 1378.50 | 99.39 | 1852.54 | 355.57 | 0.82059 | -0.59468 | -0.70647 | 0.49794 | 56.76 |
| **Me** | 1.70 | 2.20 | 3.07 | 1391.63 | 122.69 | 1831.67 | 408.67 | 0.80753 | -0.61086 | -0.70423 | 0.50162 | 0.00 |
| **F** | 1.47 | 1.47 | 2.80 | 1407.80 | 155.11 | 1837.86 | 410.90 | 0.80809 | -0.60458 | -0.68210 | 0.49971 | 0.00 |
| **Cl** | 1.77 | 1.77 | 3.47 | 1403.56 | 136.98 | 1838.98 | 404.79 | 0.81116 | -0.59957 | -0.68514 | 0.50130 | 25.80 |
| **Br** | 1.92 | 1.92 | 3.77 | 1401.92 | 130.52 | 1842.55 | 407.66 | 0.81188 | -0.59658 | -0.68619 | 0.50127 | 30.30 |
| **$CF_3$** | 2.08 | 2.71 | 3.58 | 1402.74 | 230.78 | 1854.51 | 371.41 | 0.80925 | -0.58821 | -0.68729 | 0.50302 | 36.60 |
| **CN** | 1.78 | 1.78 | 4.18 | 1403.17 | 160.60 | 1847.07 | 386.94 | 0.80262 | -0.59353 | -0.68333 | 0.50810 | 29.19 |
| **$NO_2$** | 1.55 | 2.58 | 3.61 | 1395.67 | 112.67 | 1869.31 | 375.43 | 0.81941 | -0.57682 | -0.69015 | 0.50473 | 44.26 |

Table S4 : Hammett-extended parameters for the *ortho* substituent.[36]

| *R-meta* | $B_{1,m}$ | $B_{5,m}$ | *Lm* | $\sigma_m$ | $\nu_{COH,m}$ | $I_{COH,m}$ | $\nu_{C=O,m}$ | $I_{C=O,m}$ | $NBO_{C,m}$ | $NBO_{=O,m}$ | $NBO_{O,m}$ | $NBO_{H,m}$ |
|---|---|---|---|---|---|---|---|---|---|---|---|---|
| **H** | 1.00 | 1.00 | 2.09 | 0.00 | 1394.60 | 163.25 | 1847.85 | 390.68 | 0.80516 | -0.60208 | -0.70358 | 0.50087 |
| **$NMe_2$** | 2.00 | 3.28 | 4.11 | -0.16 | 1378.42 | 106.94 | 1845.61 | 398.56 | 0.80655 | -0.60340 | -0.70782 | 0.49914 |
| **$NH_2$** | 1.55 | 2.08 | 2.99 | -0.16 | 1398.20 | 172.81 | 1847.74 | 379.29 | 0.80639 | -0.60132 | -0.70622 | 0.49971 |
| **OH** | 1.52 | 2.07 | 2.86 | 0.12 | 1398.66 | 189.34 | 1849.36 | 370.78 | 0.80529 | -0.59959 | -0.70183 | 0.50059 |
| **OCH3** | 1.52 | 3.20 | 4.25 | 0.12 | 1396.67 | 216.86 | 1847.51 | 385.34 | 0.80599 | -0.60040 | -0.70218 | 0.50019 |
| ***t*Bu** | 2.92 | 3.35 | 4.36 | -0.10 | 1395.11 | 148.49 | 1845.19 | 426.62 | 0.80583 | -0.60299 | -0.70606 | 0.50005 |
| **Me** | 1.70 | 2.19 | 3.07 | -0.07 | 1395.49 | 157.59 | 1846.08 | 400.44 | 0.80602 | -0.60256 | -0.70452 | 0.49996 |
| **F** | 1.47 | 1.47 | 2.81 | 0.34 | 1397.94 | 186.91 | 1852.75 | 369.96 | 0.80394 | -0.59554 | -0.70171 | 0.50225 |
| **Cl** | 1.77 | 1.77 | 3.51 | 0.37 | 1394.82 | 180.38 | 1852.53 | 389.42 | 0.80547 | -0.59554 | -0.70118 | 0.50252 |
| **Br** | 1.92 | 1.92 | 3.81 | 0.39 | 1394.08 | 177.79 | 1851.09 | 396.40 | 0.80558 | -0.59541 | -0.70094 | 0.50265 |
| **$CF_3$** | 2.09 | 2.72 | 3.48 | 0.43 | 1408.57 | 201.35 | 1853.23 | 388.18 | 0.80364 | -0.59415 | -0.70028 | 0.50352 |
| **CN** | 1.78 | 1.78 | 4.18 | 0.56 | 1401.10 | 182.03 | 1856.38 | 395.52 | 0.80490 | -0.59114 | -0.69879 | 0.50472 |
| **$NO_2$** | 1.55 | 2.59 | 3.56 | 0.71 | 1399.96 | 179.66 | 1856.20 | 387.73 | 0.80328 | -0.59110 | -0.69693 | 0.50490 |

Table S5 : Hammett-extended parameters for the *meta* substituent.[36]

| ***R-para*** | $B_{1,p}$ | $B_{5,p}$ | $L_p$ | $\sigma_p$ | $\nu_{COH,p}$ | $I_{COH,p}$ | $\nu_{C=O,p}$ | $I_{C=O,p}$ | $NBO_{C,p}$ | $NBO_{=O,p}$ | $NBO_{O,p}$ | $NBO_{H,p}$ |
|---|---|---|---|---|---|---|---|---|---|---|---|---|
| **H** | 1.00 | 1.00 | 2.09 | 0.00 | 1394.60 | 163.25 | 1847.85 | 390.68 | 0.80516 | -0.60208 | -0.70358 | 0.50087 |
| **$NMe_2$** | 1.55 | 3.30 | 4.33 | -0.83 | 1396.43 | 313.71 | 1830.85 | 486.47 | 0.80471 | -0.62021 | -0.71045 | 0.49757 |
| **$NH_2$** | 1.55 | 2.10 | 2.98 | -0.66 | 1395.35 | 257.80 | 1834.97 | 453.68 | 0.80524 | -0.61626 | -0.70910 | 0.49865 |
| **OH** | 1.52 | 2.07 | 2.83 | -0.37 | 1396.61 | 221.99 | 1840.65 | 422.67 | 0.80599 | -0.61060 | -0.70602 | 0.50016 |
| **OCH3** | 1.52 | 3.22 | 4.25 | -0.27 | 1396.68 | 299.76 | 1839.37 | 424.99 | 0.80480 | -0.61182 | -0.70667 | 0.49976 |
| ***t*Bu** | 2.91 | 3.35 | 4.34 | -0.20 | 1395.78 | 189.95 | 1844.67 | 435.29 | 0.80613 | -0.60531 | -0.70473 | 0.49999 |
| **Me** | 1.70 | 2.20 | 3.07 | -0.17 | 1395.02 | 178.02 | 1844.66 | 419.34 | 0.78588 | -0.60618 | -0.70599 | 0.50033 |
| **F** | 1.47 | 1.47 | 2.80 | 0.06 | 1396.19 | 204.97 | 1847.75 | 398.96 | 0.80562 | -0.60242 | -0.70414 | 0.50193 |
| **Cl** | 1.77 | 1.77 | 3.47 | 0.23 | 1395.38 | 186.22 | 1849.16 | 414.28 | 0.80492 | -0.59926 | -0.70285 | 0.50238 |
| **Br** | 1.92 | 1.92 | 3.77 | 0.23 | 1395.33 | 182.92 | 1849.58 | 419.46 | 0.80491 | -0.59850 | -0.70255 | 0.50251 |
| **$CF_3$** | 2.08 | 2.71 | 3.58 | 0.54 | 1401.11 | 45.48 | 1854.66 | 375.38 | 0.78303 | -0.59324 | -0.70152 | 0.50390 |
| **CN** | 1.78 | 1.78 | 4.18 | 0.66 | 1397.32 | 181.63 | 1855.52 | 393.01 | 0.80177 | -0.58965 | -0.69956 | 0.50474 |
| **$NO_2$** | 1.55 | 2.58 | 3.61 | 0.78 | 1396.61 | 202.70 | 1857.56 | 370.43 | 0.78101 | -0.58788 | -0.69977 | 0.50541 |

Table S6 : Hammett-extended parameters for the *para* substituent.[36]

Hammett-extended descriptors cannot account for differences in FIA values for molecules exhibiting the same substituents but with different molecular scaffolds (Figure S10).

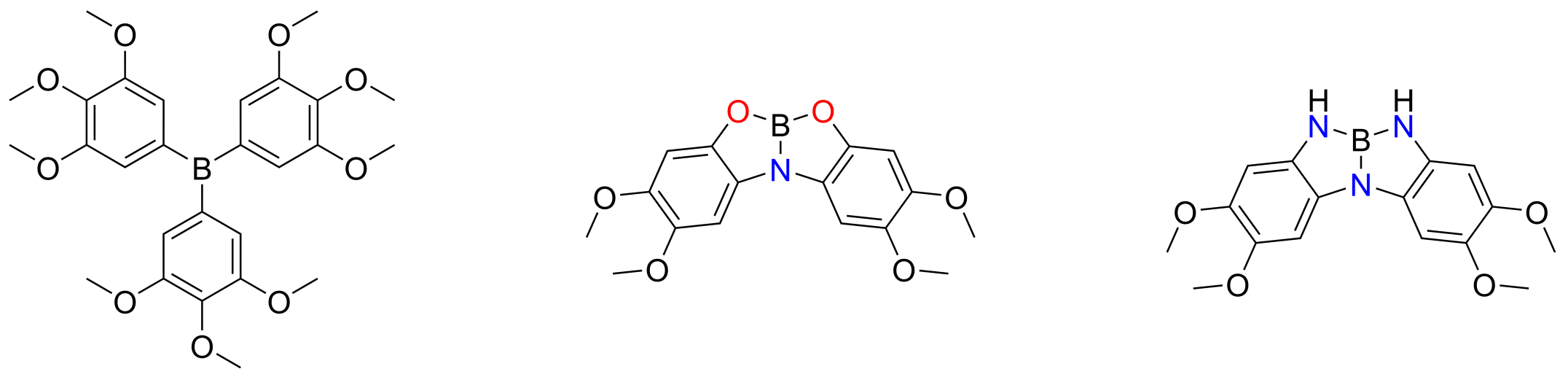

FIA ($kJ.mol^{-1}$) 343 339 273

Figure S10 : Varying FIA values for molecules with the same substituents but differing molecular scaffolds.

# S5 Constructing machine learning models

## Dataset splitting

To address the low-data regime, stratified sampling was used to split the dataset into training and testing sets, preserving the FIA distribution across both. Classes defined in the **Interpretability**, *Molecular Design* – Interpretable ML section of the manuscript for decision tree fitting were used for stratification, ensuring the testing set was representative of the overall FIA population for the ONO scaffold (Figure S11).

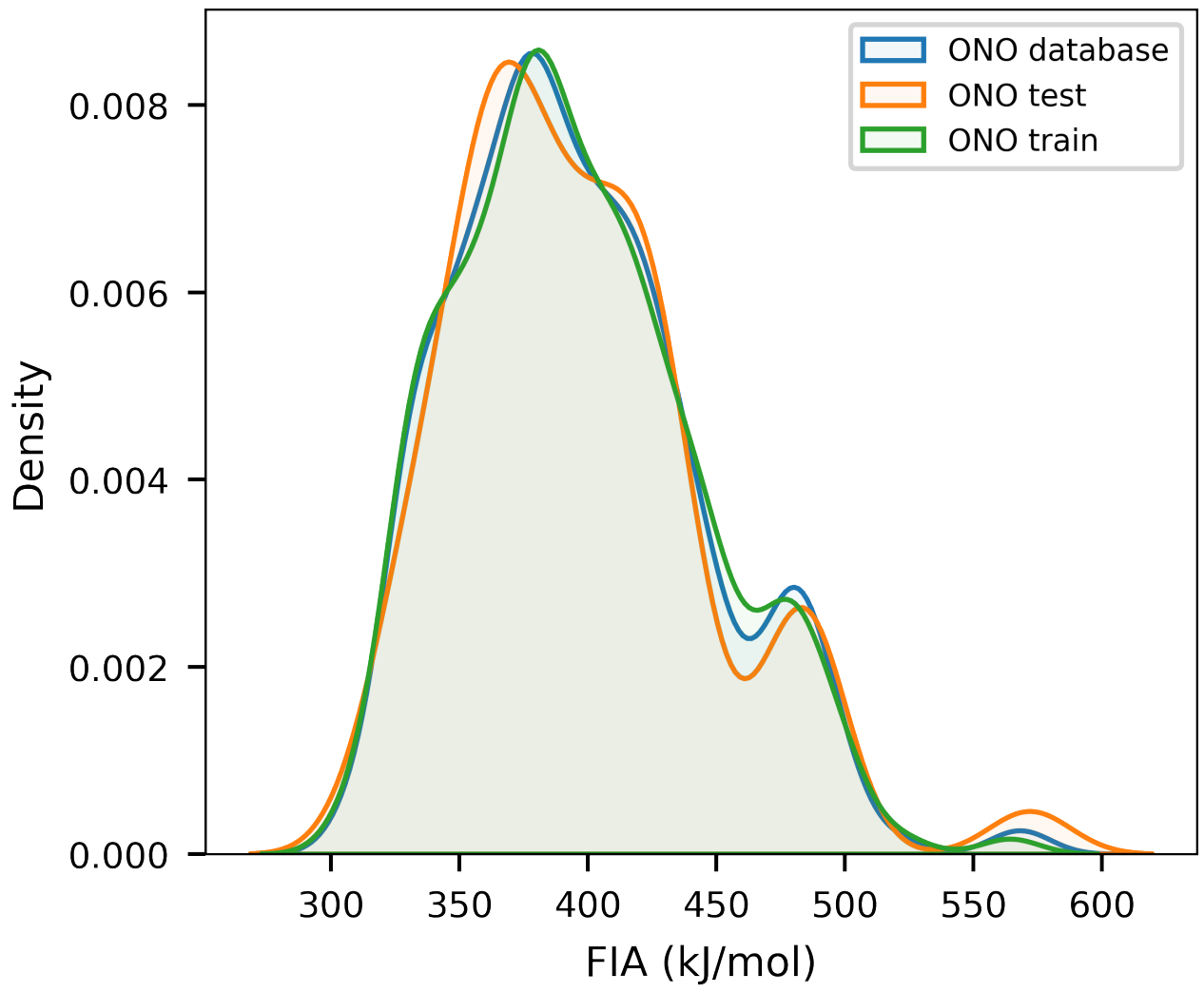

Figure S11 : FIA distribution for the whole database, for training and for testing set defined for model performances evaluation.

## Models' evaluation

### *Data preprocessing*

Before any learning (including folds of cross validation), data was preprocessed by removing constant features, fitting a standard scaler on the training set and applying it on both the training and validation sets. This way, all features were scaled to have 0 mean and unit standard deviation. This scaling process was not applied to Morgan fingerprints, as they are bit-vectors. For fingerprints, only constant features across datapoint molecules (0 variance) were removed.

The mean absolute error (MAE) was chosen as the scoring function to evaluate the models, as it provided the advantage to possess a unit, which makes it easier to Figure out what physically means the error in a regression task. Models were evaluated and optimized on the training set. The hyperparameters of the models were tuned beginning with a randomized search and then with a grid search 4-fold cross-validation scheme. Final tuning of the hyperparameters was performed manually using a 10-fold cross-validation scheme. That scheme was repeated 10 times to statistically evaluate the distribution of errors for each model.

The obtained models were used to conduct the rest of this study. Chosen hyperparameters are available in the GitHub repository.[35]

## Oracle development

Linear regression (LR) was chosen as the ML algorithm due to its strong performance across a broad range of molecular descriptors. Concatenated features from the Hammett-extended and RDKit descriptors were filtered using Scikit-learn's **SelectKBest** feature selection algorithm with **f_regression** as the scoring function.[37] This function evaluates features based on their F-statistic, measuring the strength of their linear relationship with the target FIA when considered independently. This approach retained only the most relevant features for the regression task, with the optimal number of features determined through a grid search involving also the model's hyperparameters to minimize the MAE. Notably, features such as structural fragment descriptors from RDKit and Sterimol parameters like $B_1$ and $B_5$ from the Hammett-extended descriptors were excluded due to their low correlation with FIA. See the GitHub repository for details.[35]

## Extrapolation

### *Quantum descriptors*

| Linear | LR | Bayes. Ridge | LASSO | SVR | Tree | RF | Grad. Boost. | GPR | KNN | MLP |
|---|---|---|---|---|---|---|---|---|---|---|
| MAE = 54.1 | MAE = 187.8 | MAE = 227.3 | MAE = 163.6 | MAE = 86.5 | MAE = 76.3 | MAE = 71.2 | MAE = 61.9 | MAE = 86.4 | MAE = 60.2 | MAE = 24.1 |
| a = 0.89 | a = 0.88 | a = 0.88 | a = 0.88 | a = 0.86 | a = 0.81 | a = 0.84 | a = 0.76 | a = 0.0 | a = 0.45 | a = 0.89 |
| b = -18.8 | b = 226.6 | b = 266.5 | b = 200.9 | b = 130.1 | b = 135.1 | b = 121.5 | b = 137.3 | b = 396.7 | b = 231.3 | b = 13.5 |

Table S7 : Performances of various ML algorithms trained on the ONO dataset and tested on the NNN dataset with the quantum descriptors.

| | ONO | NNN | OCO | Triarylboranes |
|---|---|---|---|---|
| **ONO** | MAE = 8.13<br>a = 0.95<br>b = 19.03 | MAE = 692.9<br>a = 0.78<br>b = 780.95 | MAE = 104.03<br>a = 0.86<br>b = 158.01 | MAE = 268.7<br>a = 1.41<br>b = -430.62 |
| **NNN** | MAE = 187.79<br>a = 0.88<br>b = 226.62 | MAE = 9.56<br>a = 0.95<br>b = 16.76 | MAE = 283.4<br>a = 0.88<br>b = -244.77 | MAE = 164.47<br>a = 1.4<br>b = -290.21 |
| **OCO** | MAE = 35.09<br>a = 0.92<br>b = -8.08 | MAE = 493.86<br>a = 0.76<br>b = 574.82 | MAE = 11.21<br>a = 0.89<br>b = 39.83 | MAE = 114.52<br>a = 0.99<br>b = -111.94 |
| **triarylboranes** | MAE = 728.52<br>a = 0.93<br>b = -696.08 | MAE = 378.86<br>a = 0.38<br>b = -135.1 | MAE = 417.73<br>a = 0.31<br>b = -140.58 | MAE = 15.85<br>a = 0.92<br>b = 30.67 |

Table S8 : MAE and fit coefficients for training scaffolds (vertical) and testing scaffolds (horizontal) for the LR regressor with the quantum descriptors (no feature selection). MAE obtained for prediction within the same scaffold were averaged over a 10-fold cross-validation scheme, repeated 5 times.

| | NNN | ONO | deviation |
|---|---|---|---|
| **X** | 0.000 | -0.001 | 1.424 |
| **lumo_energy** | -0.002 | -0.010 | 1.239 |
| **Z** | -0.010 | -0.003 | 1.031 |
| **NPA_Rydberg** | 0.016 | 0.029 | 0.575 |
| **ES_root_NPA_Rydberg** | 0.017 | 0.030 | 0.564 |
| **NMR_anisotropy** | 16.345 | 11.588 | 0.341 |
| **dipole** | 4.841 | 3.584 | 0.298 |
| **Y** | -0.538 | -0.689 | 0.247 |
| **ES_root_NPA_charge** | 1.045 | 1.281 | 0.203 |
| **NPA_charge** | 1.048 | 1.258 | 0.182 |
| **APT_charge** | 1.067 | 1.254 | 0.161 |
| **ES_root_dipole** | 4.674 | 3.996 | 0.156 |
| **ES_root_NPA_valence** | 1.940 | 1.691 | 0.137 |
| **NPA_valence** | 1.937 | 1.714 | 0.122 |
| **G_thermal_correction** | 0.255 | 0.231 | 0.1 |
| **zero_point_correction** | 0.308 | 0.283 | 0.086 |
| **E_thermal_correction** | 0.331 | 0.305 | 0.083 |
| **H_thermal_correction** | 0.332 | 0.306 | 0.083 |
| **electronegativity** | 0.127 | 0.138 | 0.08 |
| **ES_root_NPA_total** | 3.955 | 3.719 | 0.062 |
| **number_of_atoms** | 40.918 | 38.485 | 0.061 |
| **NPA_total** | 3.952 | 3.742 | 0.055 |
| **homo_energy** | -0.252 | -0.265 | 0.051 |
| **H** | -2440.757 | -2541.185 | 0.04 |
| **electronic_spatial_extent** | 13522.813 | 12986.665 | 0.04 |
| **E_scf** | -2440.941 | -2541.330 | 0.04 |
| **E_zpe** | -2440.781 | -2541.208 | 0.04 |
| **G** | -2440.833 | -2541.260 | 0.04 |
| **E** | -2440.758 | -2541.186 | 0.04 |
| **NMR_shift** | 89.842 | 86.465 | 0.038 |
| **ES_root_molar_volume** | 2671.401 | 2588.301 | 0.032 |
| **molar_volume** | 2688.979 | 2605.690 | 0.031 |
| **ES_root_electronic_spatial_extent** | 13768.427 | 13392.656 | 0.028 |
| **VBur** | 0.605 | 0.590 | 0.025 |
| **hardness** | 0.125 | 0.127 | 0.021 |
| **ES_root_Mulliken_charge** | 0.670 | 0.683 | 0.02 |
| **molar_mass** | 395.729 | 390.020 | 0.015 |
| **Mulliken_charge** | 0.668 | 0.667 | 0.001 |
| **converged** | 1.000 | 1.000 | 0 |
| **ES_root_NPA_core** | 1.999 | 1.999 | 0 |
| **NPA_core** | 1.999 | 1.999 | 0 |
| **multiplicity** | 1.000 | 1.000 | 0 |

Table S9 : mean of quantum features for ONO and NNN scaffolds, and the deviation between the two scaffolds. This deviation was calculated by the formula: $\text{deviation}_{\text{feature}} = (\text{mean}_{\text{feat., NNN}}) - \text{mean}_{\text{feat.,ONO}})/(\text{mean}_{\text{feat.,NNN}} + \text{mean}_{\text{feat.,ONO}})$.

| features | corr with FIA |
|---|---|
| **homo_energy** | 0.749 |
| **electronegativity** | 0.704 |
| **lumo_energy** | 0.628 |
| **G_thermal_correction** | 0.448 |
| **zero_point_correction** | 0.403 |
| **G** | 0.378 |
| **H** | 0.378 |
| **E** | 0.378 |
| **E_zpe** | 0.378 |
| **E_scf** | 0.378 |
| **H_thermal_correction** | 0.375 |
| **E_thermal_correction** | 0.375 |
| **NPA_total** | 0.326 |
| **NPA_charge** | 0.326 |
| **NPA_valence** | 0.324 |
| **molar_mass** | 0.306 |
| **Mulliken_charge** | 0.277 |
| **ES_root_electronic_spatial_extent** | 0.271 |
| **electronic_spatial_extent** | 0.27 |
| **ES_root_Mulliken_charge** | 0.237 |
| **ES_root_NPA_total** | 0.236 |
| **ES_root_NPA_charge** | 0.236 |
| **ES_root_NPA_valence** | 0.233 |
| **number_of_atoms** | 0.23 |
| **VBur** | 0.164 |
| **APT_charge** | 0.158 |
| **NMR_anisotropy** | 0.129 |
| **dipole** | 0.127 |
| **NMR_shift** | 0.092 |
| **ES_root_NPA_Rydberg** | 0.09 |
| **NPA_Rydberg** | 0.076 |
| **hardness** | 0.07 |
| **ES_root_dipole** | 0.057 |
| **Z** | 0.054 |
| **NPA_core** | 0.039 |
| **ES_root_NPA_core** | 0.039 |
| **Y** | 0.031 |
| **molar_volume** | 0.022 |
| **X** | 0.009 |
| **ES_root_molar_volume** | 0.004 |
| **charge** | |
| **converged** | |
| **multiplicity** | |

Table S10 : correlation of quantum features with FIA (Pearson's correlation coefficient), total chemical space (ONO, NNN, OCO and triarylboranes).

*<u>Details of feature selection</u>*

The selected quantum features for the prediction from the ONO scaffold to the NNN scaffold are : E, ES_root_electronic_spatial_extent, ES_root_molar_volume, E_scf, E_thermal_correction, E_zpe, G, G_thermal_correction, H, H_thermal_correction, electronegativity, electronic_spatial_extent, hardness, homo_energy, molar_mass, molar_volume, number_of_atoms, zero_point_correction, APT_charge, ES_root_Mulliken_charge,

ES_root_NPA_charge, ES_root_NPA_core, ES_root_NPA_total, ES_root_NPA_valence, Mulliken_charge, NMR_anisotropy, NMR_shift, NPA_charge, NPA_core, NPA_total, NPA_valence.

*Exploration of the inherent differences between scaffolds*

Even when using all quantum features, models exhibited different biases depending on the structure (Table S8). This variability may stem from inherent differences between molecular scaffolds, not only in terms of quantum features but also in FIA distribution (Figure 2.B). Indeed, most ML algorithms perform well when the distribution of the training set resembles that of the testing set. Expecting the model would capture the differences between scaffolds, we trained it on three molecular scaffolds and then tested on the fourth. This approach resulted in improved performances on ONO and NNN (MAE decreased of 50 to 100 kJ.mol$^{-1}$), but not on the other structures (Table S11). The low prediction performance on triarylboranes is expected (MAE = 466 kJ.mol$^{-1}$), given their structural differences from the others, being non planar and lacking heteroatoms bonded to the boron atom (Figure 6). Furthermore, while the FIA range for the OCO structure falls within the overlap of distributions for the other structures, its FIA distribution significantly differs due to its sharpness, possibly explaining the low prediction performance (MAE = 157 kJ.mol$^{-1}$). This sharp distribution may also disrupt FIA learning when predicting on other structures. Indeed, excluding the OCO structure significantly improved performances on ONO (MAE = 25.2 kJ.mol$^{-1}$) and NNN (MAE = 12.87 kJ.mol$^{-1}$) (Table S12).

| | ONO | NNN | OCO | Triarylboranes |
|---|---|---|---|---|
| **Performance** | MAE = 73.23<br>a = 1.11<br>b = -117.61 | MAE = 41.13<br>a = 1.05<br>b = 24.72 | MAE = 157.11<br>a = 0.85<br>b = 209.43 | MAE = 466.06<br>a = 0.79<br>b = -382.58 |

Table S11 : Performances of LR and quantum descriptors model, tested on one structure while trained on the three others.

| | ONO | NNN | Triarylboranes |
|---|---|---|---|
| **Performance** | MAE = 25.21<br>a = 1.18<br>b = -94.64 | MAE = 12.87<br>a = 0.95<br>b = 21.21 | MAE = 576.75<br>a = 0.78<br>b = -490.18 |

Table S12 : Performances of LR and quantum descriptors model, tested on one structure while trained on the two others (OCO structure excluded).

| | ONO | NNN | OCO |
|---|---|---|---|
| **Performance** | MAE = 42.54<br>a = 0.92<br>b = -11.73 | MAE = 83.5<br>a = 0.9<br>b = -52.42 | MAE = 45.28<br>a = 0.91<br>b = -13.14 |

Table S13 : Performances of LR and quantum descriptors model, tested on one structure while trained on the two others (triarylboranes excluded).

### *RDKit descriptors*

Given the complexity of the RDKit descriptors, we implemented a slightly different strategy for extrapolating. We again selected the LR algorithm, as it demonstrates strong performance, achieving a MAE of 6.93 kJ.mol$^{-1}$ and 10.5 kJ.mol$^{-1}$ for the ONO and NNN respectively (Table S14). And again, when trying to predict on the NNN while trained on the ONO, the MAE of the model significantly increases to 1828 kJ.mol$^{-1}$. We then used a recursive feature elimination strategy, by iteratively removing features when it improved model predictions from ONO to NNN. Removing FractionCSP3, BCUT2D_CHGHI, and MinEStateInde really improved the performance of the model (MAE dropped to 19.97 kJ.mol$^{-1}$). Eliminating SlogP_VSA2, the MAE could be reduced to 9.05 kJ.mol$^{-1}$,while removing only 4 features out of 208. Prediction performance remained high within the ONO (7.11 kJ.mol$^{-1}$) and the NNN (10.6 kJ.mol$^{-1}$) chemical spaces using these selected features. Results are illustrated in Figure S12.

Despite these improvements, when attempting to extrapolate to other molecular scaffolds using the same selected features, the MAE remains considerably large (267.77 kJ.mol$^{-1}$ for OCO and 384.78 kJ.mol$^{-1}$ for triarylboranes), although notably reduced with respect to the initial values (Table S14). Training on three scaffolds and testing on the remaining one while keeping all RDKit features significantly improved performances (Table S15). This strategy, combined with the selected RDKit features, reduced MAE of 10 to 30 kJ.mol$^{-1}$ (Table S16).

| | **ONO** | **NNN** | **OCO** | **triarylboranes** |
|---|---|---|---|---|
| **ONO** | MAE = 6.93<br>a = 0.97<br>b = 13.6 | MAE = 4264.81<br>a = 1.0<br>b = 4263.01 | MAE = 3192.32<br>a = 0.61<br>b = 3345.35 | MAE = 126.18<br>a = 0.74<br>b = 229.8 |
| **NNN** | MAE = 1827.67<br>a = 0.86<br>b = -1782.21 | MAE = 10.46<br>a = 0.95<br>b = 15.98 | MAE = 3990.9<br>a = 0.56<br>b = -3851.26 | MAE = 141.15<br>a = 0.62<br>b = 259.66 |
| **OCO** | MAE = 1011.51<br>a = 1.04<br>b = -1023.54 | MAE = 2209.3<br>a = 0.97<br>b = 2218.55 | MAE = 9.8<br>a = 0.89<br>b = 36.66 | MAE = 130.61<br>a = 0.88<br>b = 173.01 |
| **Triarylboranes** | MAE = 3477.93<br>a = 0.68<br>b = -3354.13 | MAE = 4909.6<br>a = 9.65<br>b = -8256.87 | MAE = 9326.99<br>a = -0.73<br>b = -8656.36 | MAE = 28.94<br>a = 0.81<br>b = 74.53 |

Table S14 : MAE and fit coefficients for training scaffolds (vertical) and testing scaffolds (horizontal) for the LR regressor with non-selected RDKit features model (10-fold cross-validation scheme).

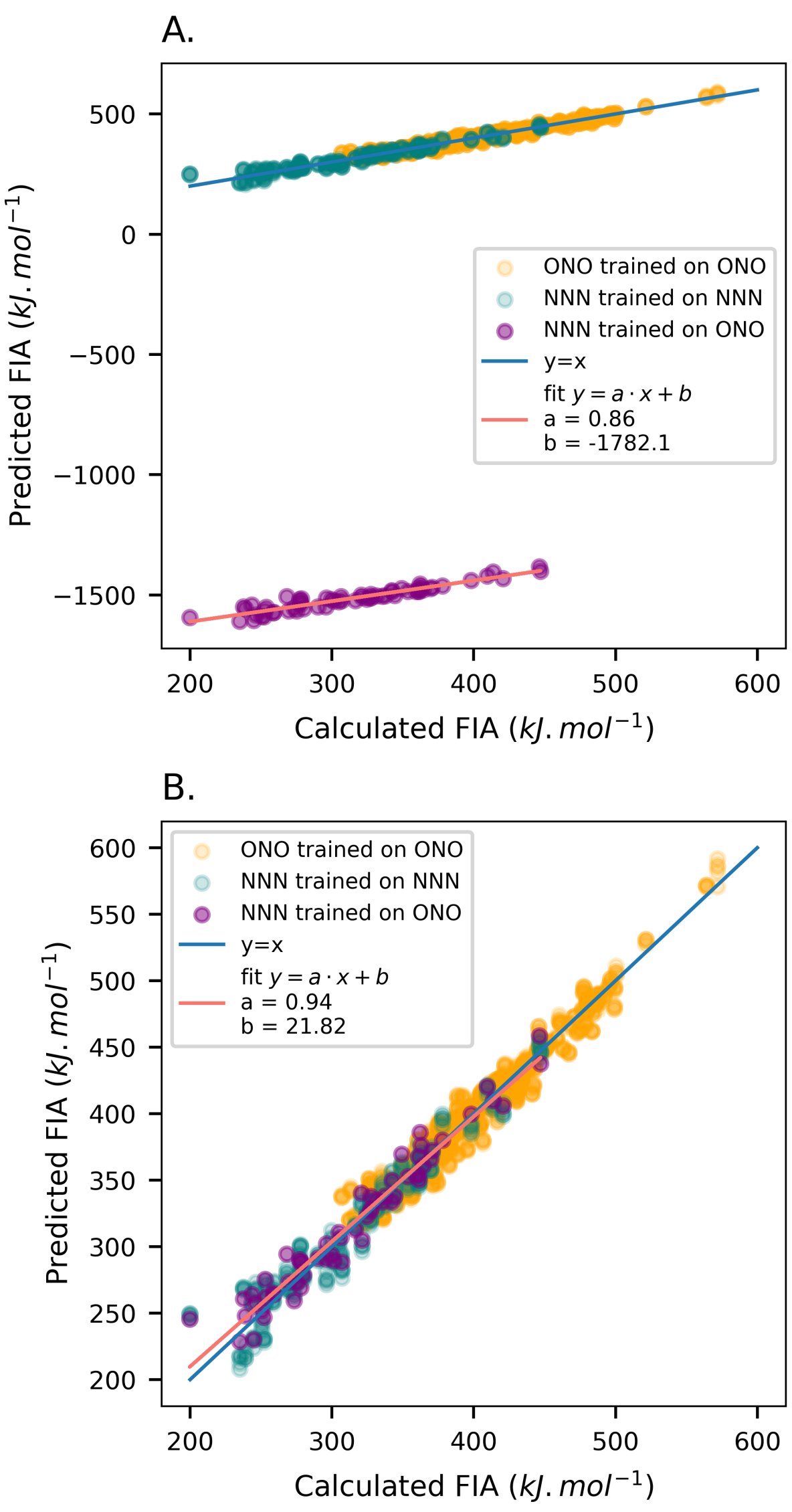

Figure S12 : Feature selection to extrapolate from ONO to NNN with the LR regressor and RDKit descriptors. A. No feature selection. B. Features selected.

| | ONO | NNN | OCO | Triarylboranes |
|---|---|---|---|---|
| **Performance** | MAE = 37.7<br>a = 0.81<br>b = 37.66 | MAE = 56.25<br>a = 0.72<br>b = 145.36 | MAE = 56.42<br>a = 0.98<br>b = 62.64 | MAE = 87.41<br>a = 1.08<br>b = -93.03 |

Table S15 : Performances of LR and non-selected RDKit descriptors model, tested on one structure while trained on the three others.

| | ONO | NNN | OCO | Triarylboranes |
|---|---|---|---|---|
| **Performance** | MAE = 21.49<br>a = 0.77<br>b = 72.2 | MAE = 45.33<br>a = 0.73<br>b = 131.67 | MAE = 25.06<br>a = 0.98<br>b = 28.95 | MAE = 79.38<br>a = 1.11<br>b = -88.95 |

Table S16 : Performances of LR and selected RDKit descriptors model, tested on one structure while trained on the three others.

### *Hammett-extended descriptors*

As expected for the Hammett-extended descriptors that do not take into account the molecular scaffold, no feature selection could improve the MAE of 86.4 kJ.mol$^{-1}$ to predict from ONO to NNN.

# S6 Models interpretability

## Lewis acidity interpretability

### *Decorrelating features*

Various features interpretation techniques presuppose that the features are not correlated, which is why we first searched to decorrelate features prior to any analysis. We computed the correlation matrix (using the Spearman correlation coefficient) and performed hierarchical clustering to identify uncorrelated features. The threshold to separate clusters was set to 0.45 (Figure S13.A). Then we chose one feature by cluster (the easiest to interpret physically), except for boron atom coordinates for which we chose only the X coordinate. This resulted in 12 chosen uncorrelated features (Figure S13.B).

Most models, except the tree-based ensemble models, were less performant on uncorrelated features (Table S17 & Table S18). Yet, this is less crucial for our purposes as reliable explanations can be derived from reasonably good predictive models.

### *Interpretability*

| | **Linear** | **LR** | **Bayes. Ridge** | **LASSO** | **SVR** | **Tree** | **RF** | **Grad. Boost.** | **GPR** | **KNN** | **MLP** |
|---|---|---|---|---|---|---|---|---|---|---|---|
| **All features** | 16.27 | 18.09 | 16.82 | 16.18 | 18.16 | 21.53 | 14.58 | 13.39 | 17.14 | 20.57 | 16.1 |
| **Uncorrelated features** | 20.93 | 21.21 | 21.01 | 20.96 | 21.8 | 20.69 | 15.7 | 14.61 | 21.22 | 21.43 | 20.94 |

Table S17 : MAEs for selected models evaluated on all database (ONO, NNN, OCO, triarylboranes) with the quantum descriptors reduced or not to uncorrelated features (10-fold cross validation repeated 10 times with different split).

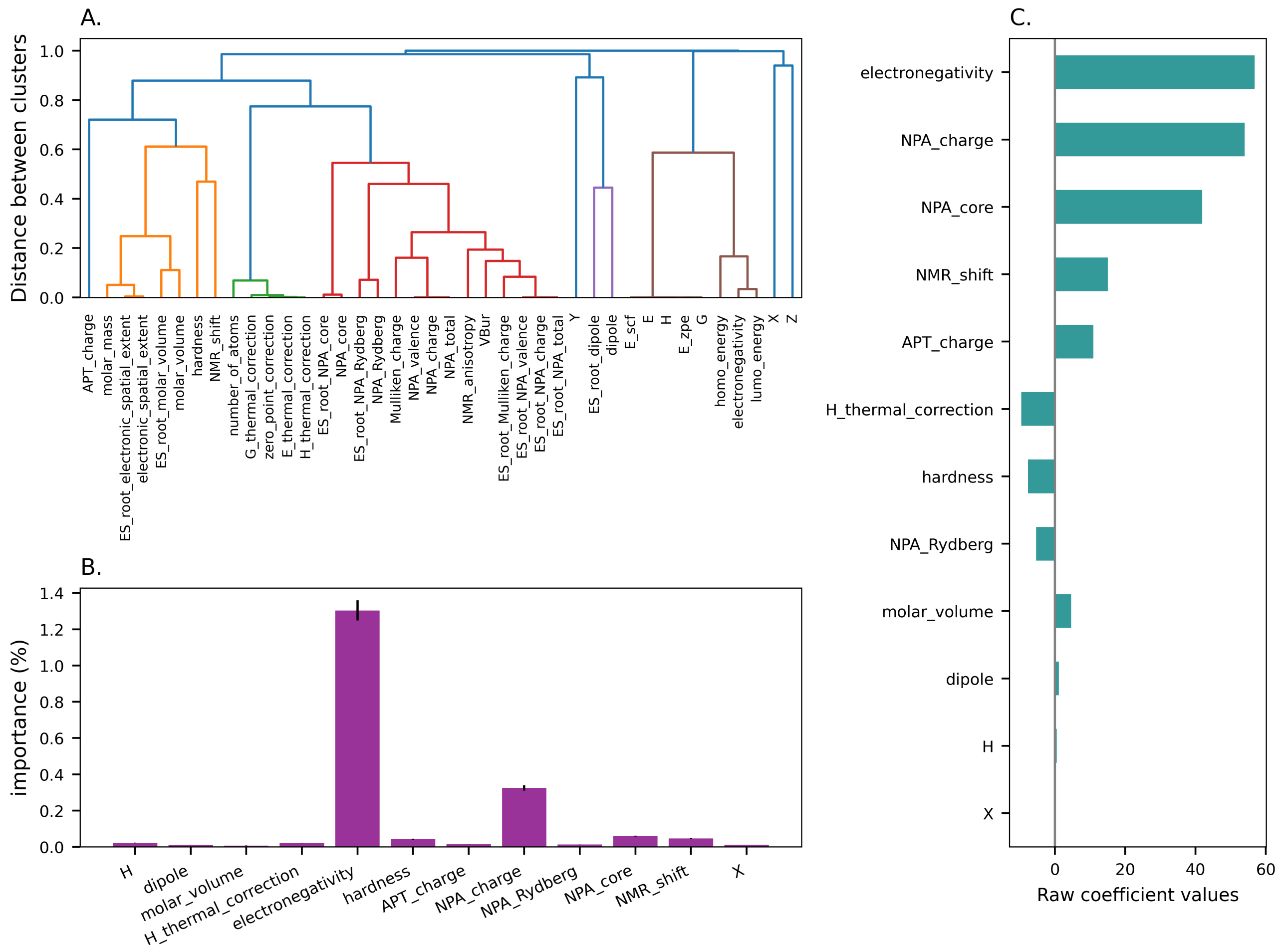

Figure S13 : Interpretability of the quantum features. Models fitted on the whole database (ONO, NNN, OCO, triarylboranes). A. Hierarchical clustering on quantum features. B. Permutation of feature importance for the gradient boosting regressor. C. Linear regression model coefficients.

The high coefficient of core electrons (NPA_core) in the linear regression model is likely an artifact (Figure S13), as it strongly correlates with the molecular scaffolds (Figure S14).

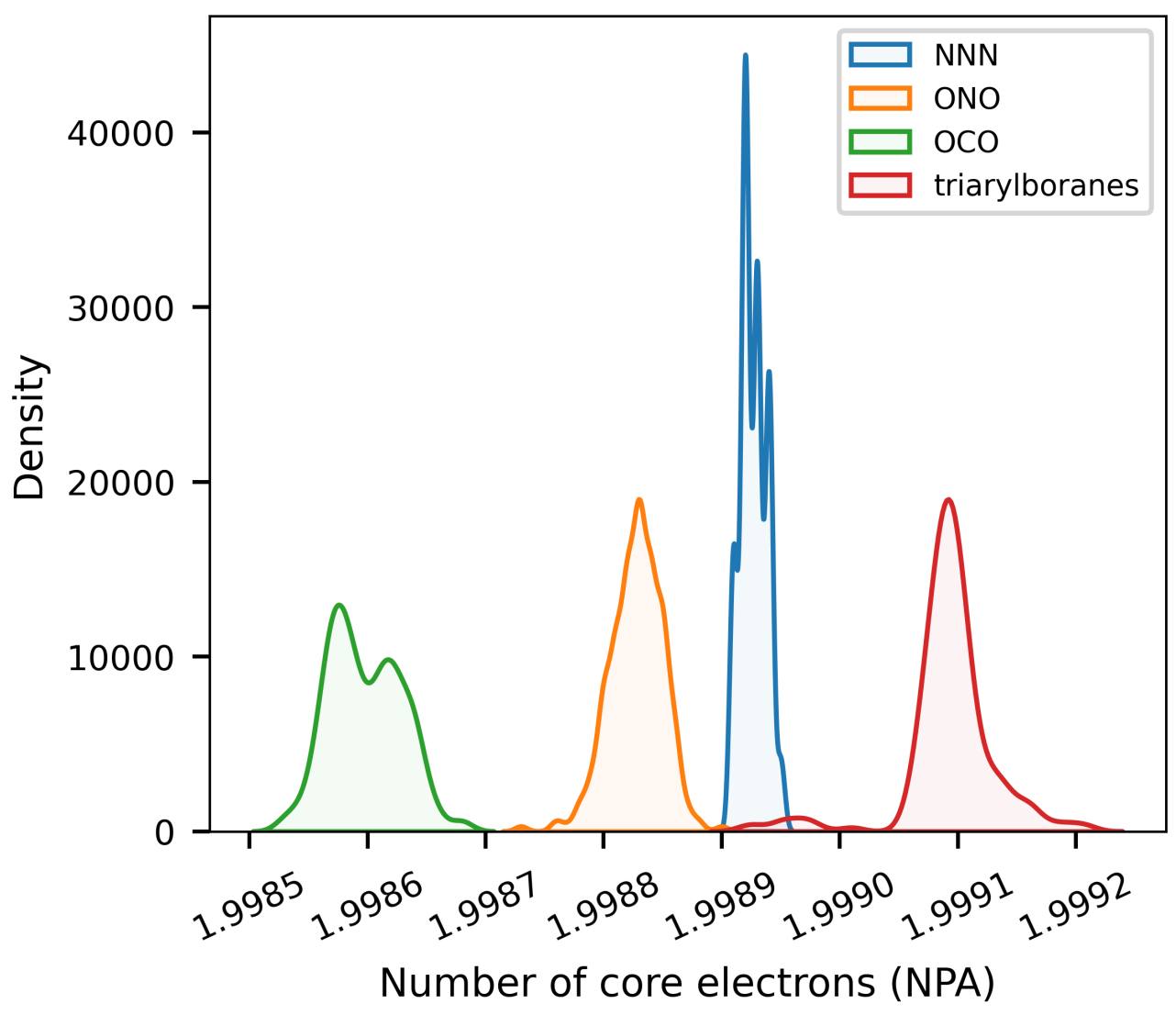

Figure S14 : Distributions of the number of core electrons among the database.

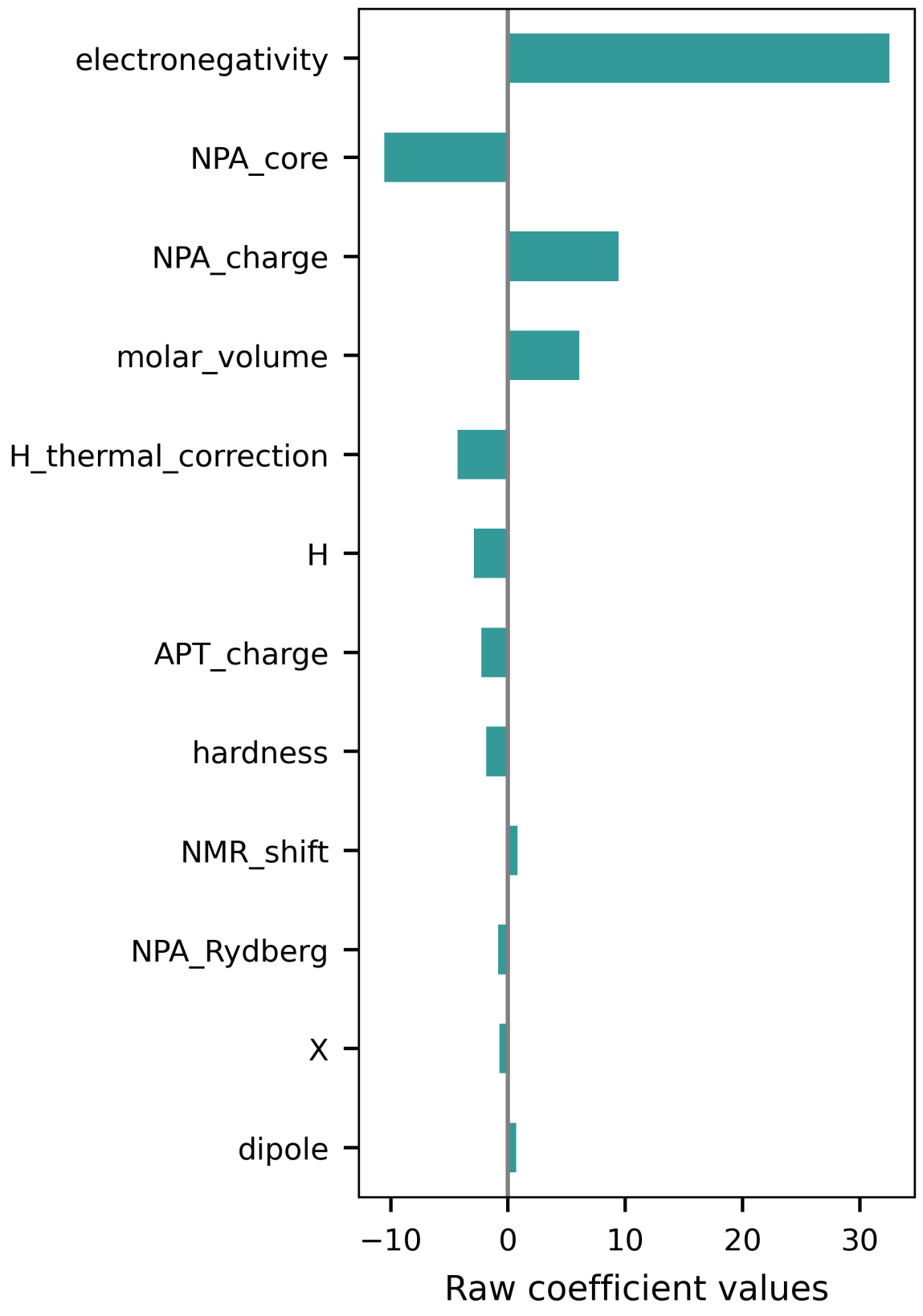

Figure S15 : Coefficients of a linear regression model fitted on uncorrelated features for the ONO molecular scaffold only.

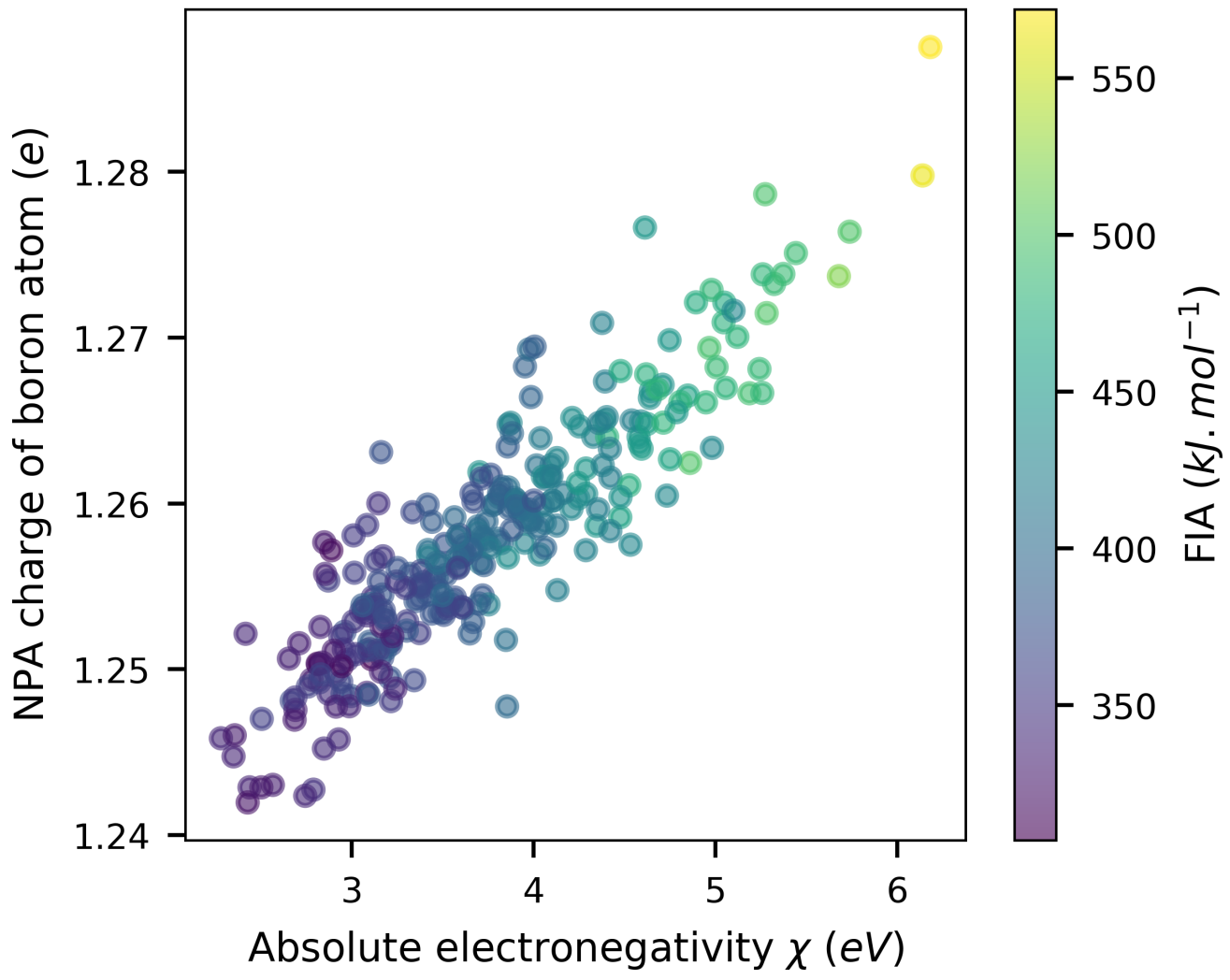

Figure S16 : FIA plot as a function of NPA charge of boron atom and absolute electronegativity of the molecule, for the ONO scaffold. *FIA = 60.0* $\chi$ *+ 8.15 NPA charge + 161*, $R^2$ = 0.88.

## Rationalization of the effect of the nature and position of substituents

### *Decorrelating features*

Hierarchical clustering was performed to decorrelate features (Figure S18 A & C). The threshold was set to 0.55 for the two scaffolds.

### *Interpretability*

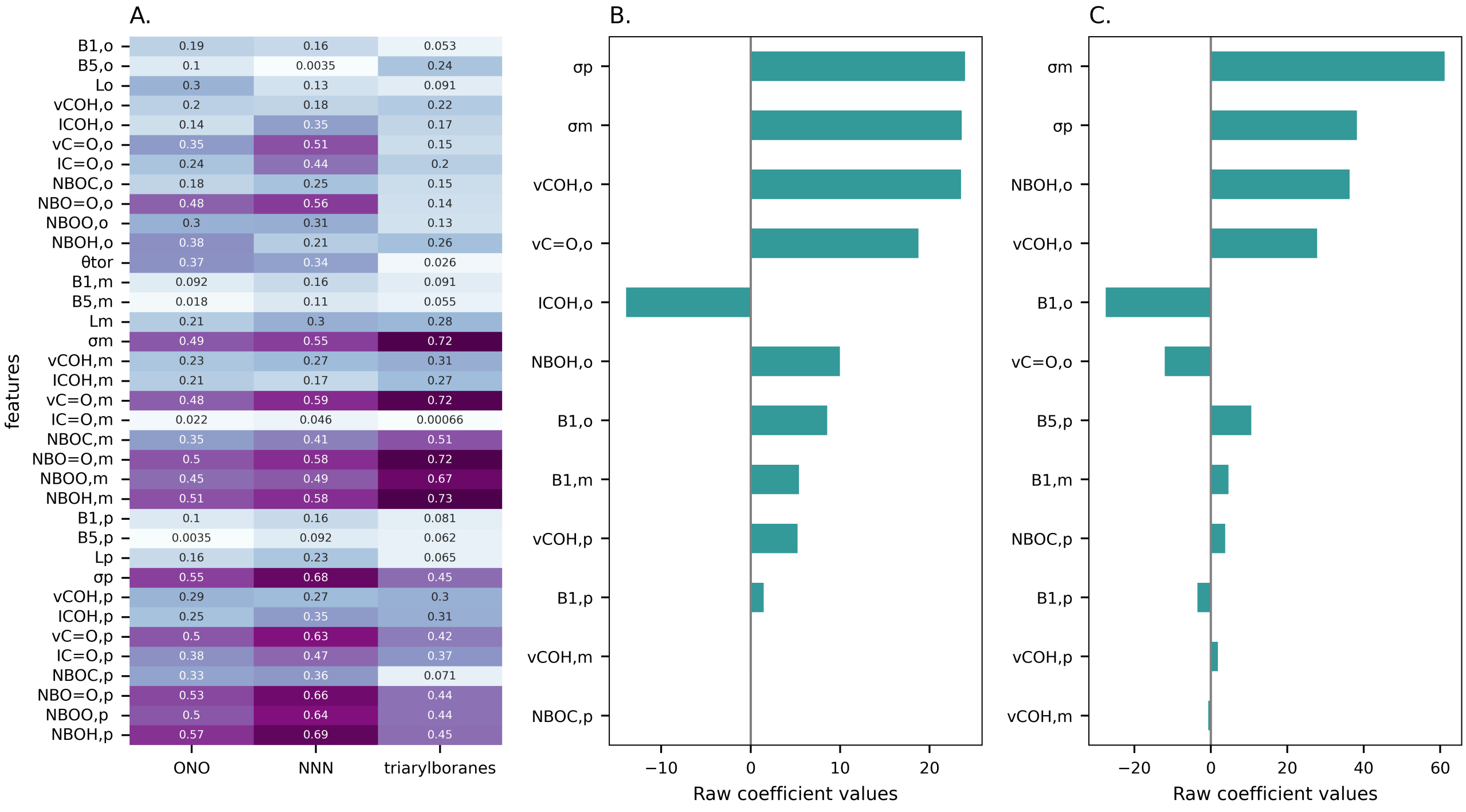

Figure S17 : Analyzing Hammett-extended descriptors. A. Pearson's correlation coefficients of features with FIA for ONO, NNN and triarylboranes molecular scaffolds. B. Coefficients of linear regression model fitted on the ONO dataset using selected uncorrelated features. C. Coefficients of linear regression model fitted on the triarylboranes dataset using selected uncorrelated features.

| | Linear | LR | Bayes. Ridge | LASSO | SVR | Tree | RF | Grad. Boost. | GPR | KNN | MLP |
|---|---|---|---|---|---|---|---|---|---|---|---|
| **ONO** | 15.27 | 15.29 | 15.24 | 15.23 | 16.1 | 14.79 | 10.56 | 6.52 | 12.07 | 18.94 | 15.29 |
| **Triarylboranes** | 22.11 | 22.14 | 22.1 | 22.12 | 25.98 | 26.87 | 21.77 | 16.66 | 29.21 | 35.21 | 22.02 |

Table S18 : MAEs ($kJ.mol^{-1}$) of models for ONO and triarylboranes with the uncorrelated Hammett-extended descriptors (10-fold cross-validation repeated 10 times).

We chose gradient boosting regressor model to perform permutation importance as it had the best performance for both scaffolds using the selected uncorrelated features.

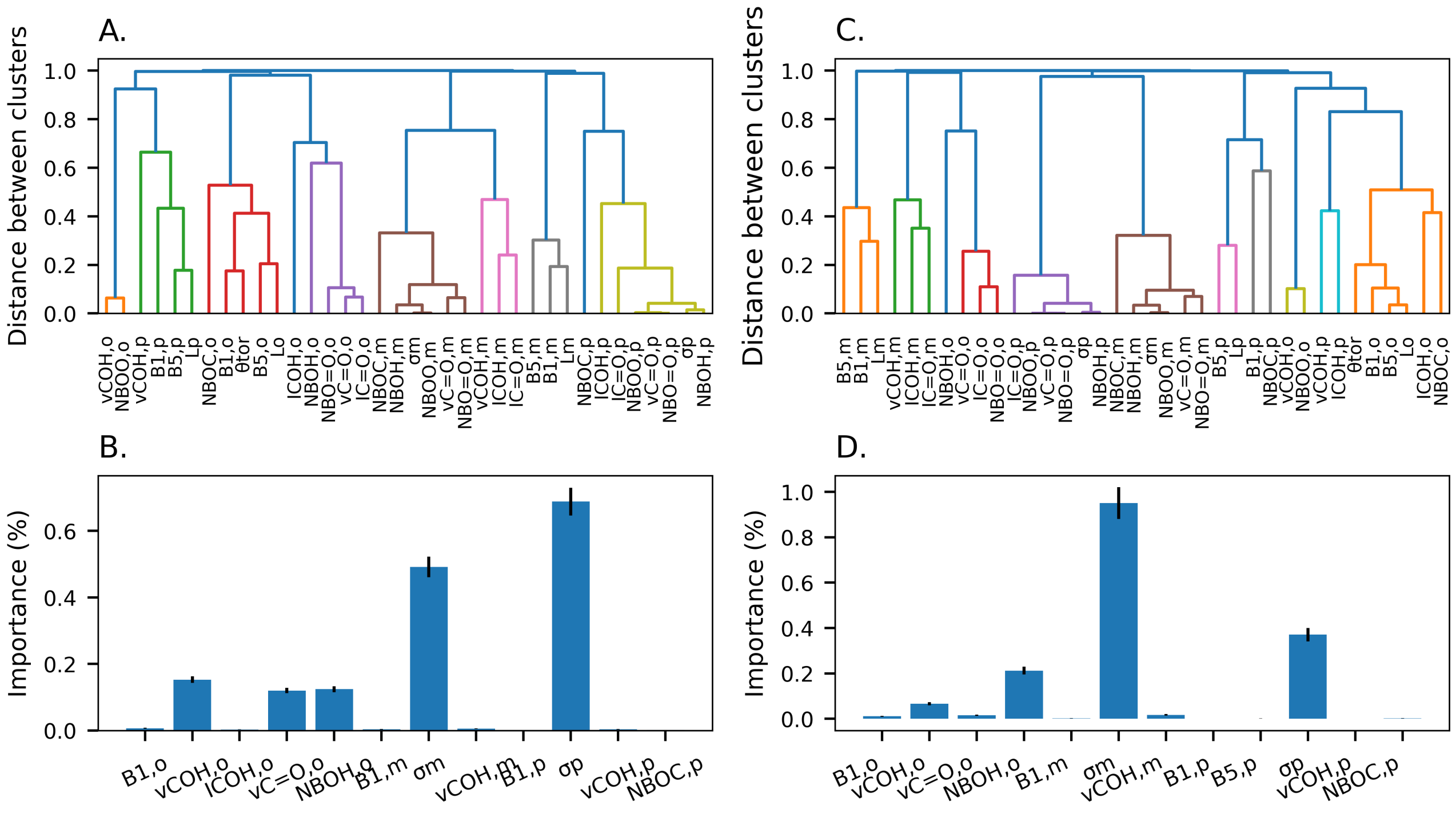

Figure S18 : Permutation importance on Hammett-extended descriptors. A & B.ONO structure. C & D. Triarylboranes.

### *Comparison between ONO and triarylboranes scaffolds*

Analysis of the coefficients of linear model and of the permutation importance algorithm reveals that for the ONO scaffold, $\sigma_p$ is slightly more important than $\sigma_m$, while it is the opposite for triarylboranes, $\sigma_m$ predominates, followed by $\sigma_p$. This indicates that when designing a triarylborane compound, fixing the *meta* group already determines the range of FIA. The final value can be refined by the choice of substituents put at the *ortho* and *para* positions.

### *Decision trees*

The data was not scaled in this part of the analysis to simplify the interpretation of the threshold values at split nodes. Moreover, decision trees are generally insensitive to the scale of the data. A leaf is considered purer (i.e., it contains entries belonging to the same class) when its impurity measure is closer to zero. For this study, models were built using selected Hammett features for simplification. The maximum depth of the trees was limited to 3 for easier interpretation. A criterion on entropy (a measure of the impurity of the leaf) was employed.

#### *ONO scaffold*

Classes are: medium LA, good LA, strong LA, super LA and selected features are: $NBO{=}O_o$, $NBO{=}O_m$, $NBO{=}O_p$, and $L_o$. The decision tree model was evaluated using 10-fold cross-validation repeated 10 times, achieving an accuracy of 0.73. This value is satisfying to derive reliable interpretations. The model was then fitted on the whole ONO dataset and plotted (Figure S19).

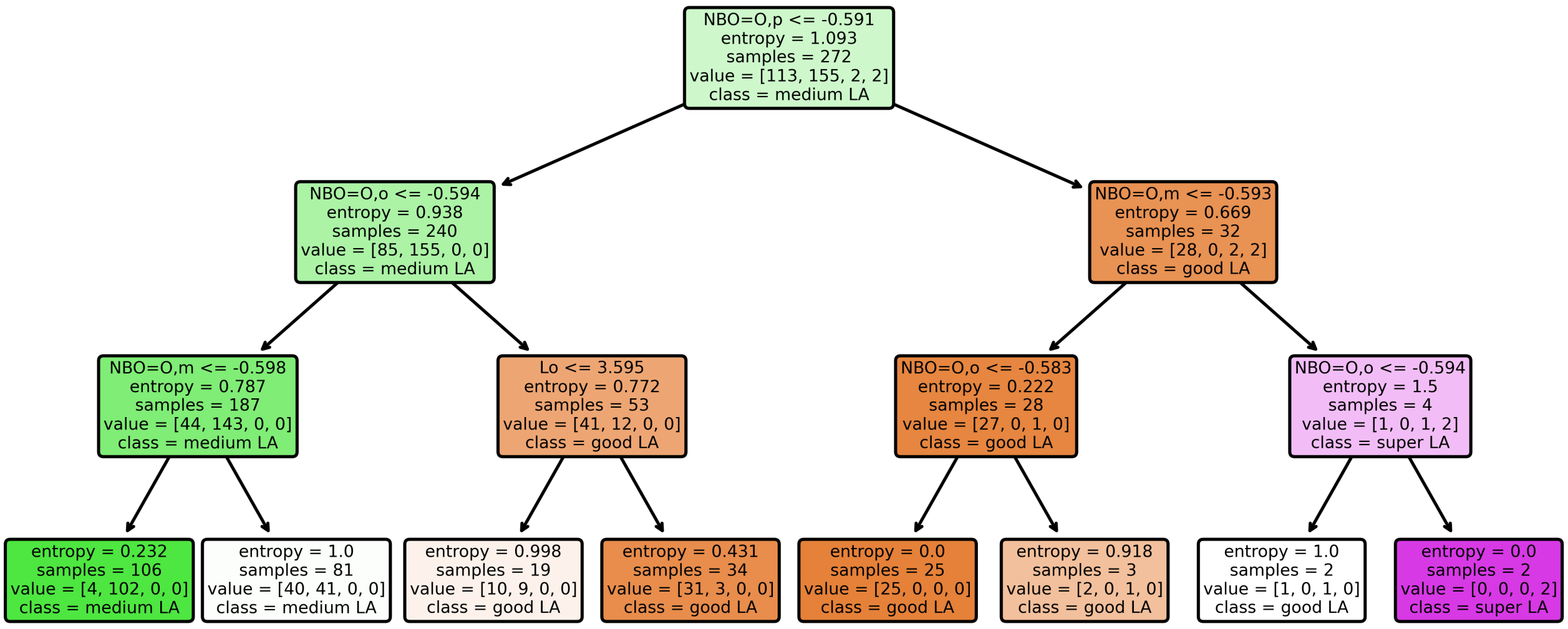

Figure S19 : Scikit-learn[37] tree fitted on the whole dataset for ONO scaffold (0.73 accuracy on 10-fold CV).

*Triarylboranes*

FIA distribution for triarylboranes being broader, classes are: very weak LA, weak LA, medium LA, good LA, strong LA, super LA. We started using $NBO_{=O}$ partial charges, torsional angle θtor and Lo as the selected features. The cross-validation accuracy score of the resulting tree is 0.48 and $NBO_{=O,o}$ and $L_o$ are not used by the model. As the electronic demand of the *ortho* substituent seemed less important, we used only $\sigma_m$, $\sigma_p$, and replaced Lo by θtor for easier interpretation (accuracy score is then 0.49 on 10-fold CV). The results were less straightforward compared to the ONO dataset. The root node assesses whether the *meta* substituent is electron-donating or not, which partly cleaves the database between good and medium LA. This finding is in agreement with the previous interpretations on Hammett-extended descriptors, namely that the *meta* substituent is the most important to determine FIA value. However, even with an electron-withdrawing group in the *meta* position, medium LA can still be obtained, provided that this group does not exhibit strong mesomeric attraction and is paired with a *para* group that acts as a good donor, such as an amine or tertiobutyl (*t*Bu). Steric effects play a pivotal role in distinguishing between molecules featuring an electron-donating group in the *meta* position and a *para* group with limited electronic attraction (e.g., not $NO_2$, CN, or $CF_3$): if the *ortho* group induces sufficient steric hindrance to cause a torsional angle θtor between the carboxylic acid moiety and the ring, the LA tends to be medium or weak; otherwise, it may still be good.

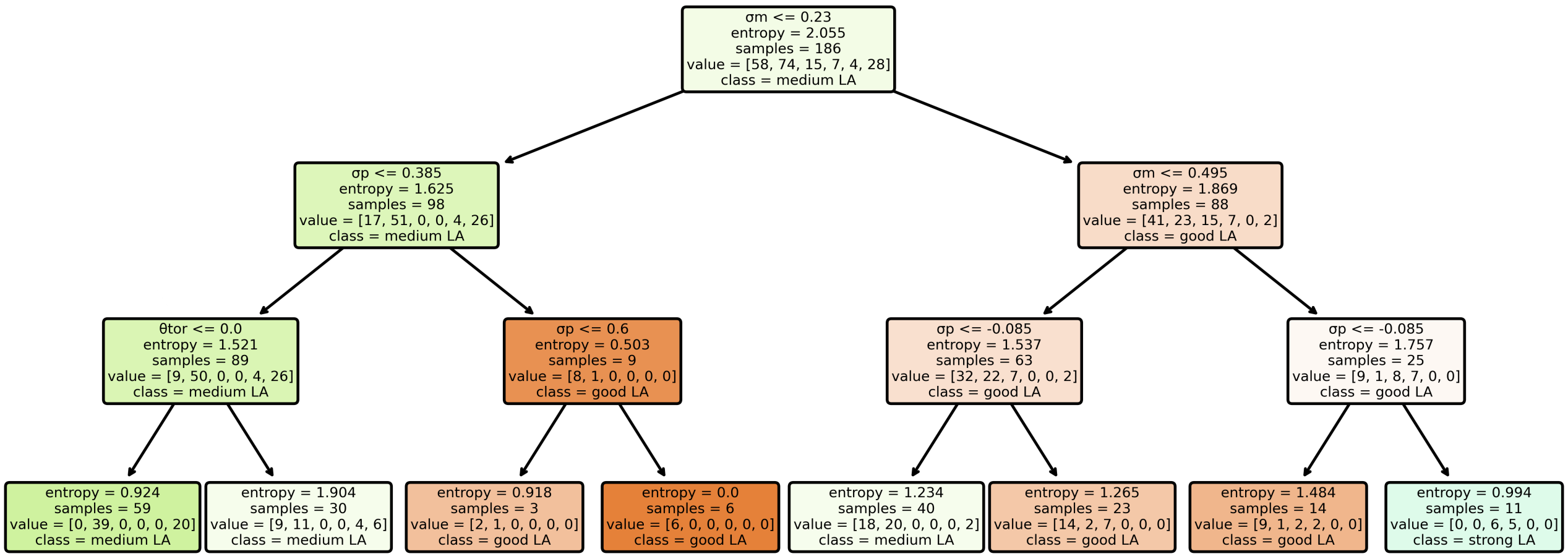

Figure S20 : Triarylboranes Scikit-learn[37] tree.

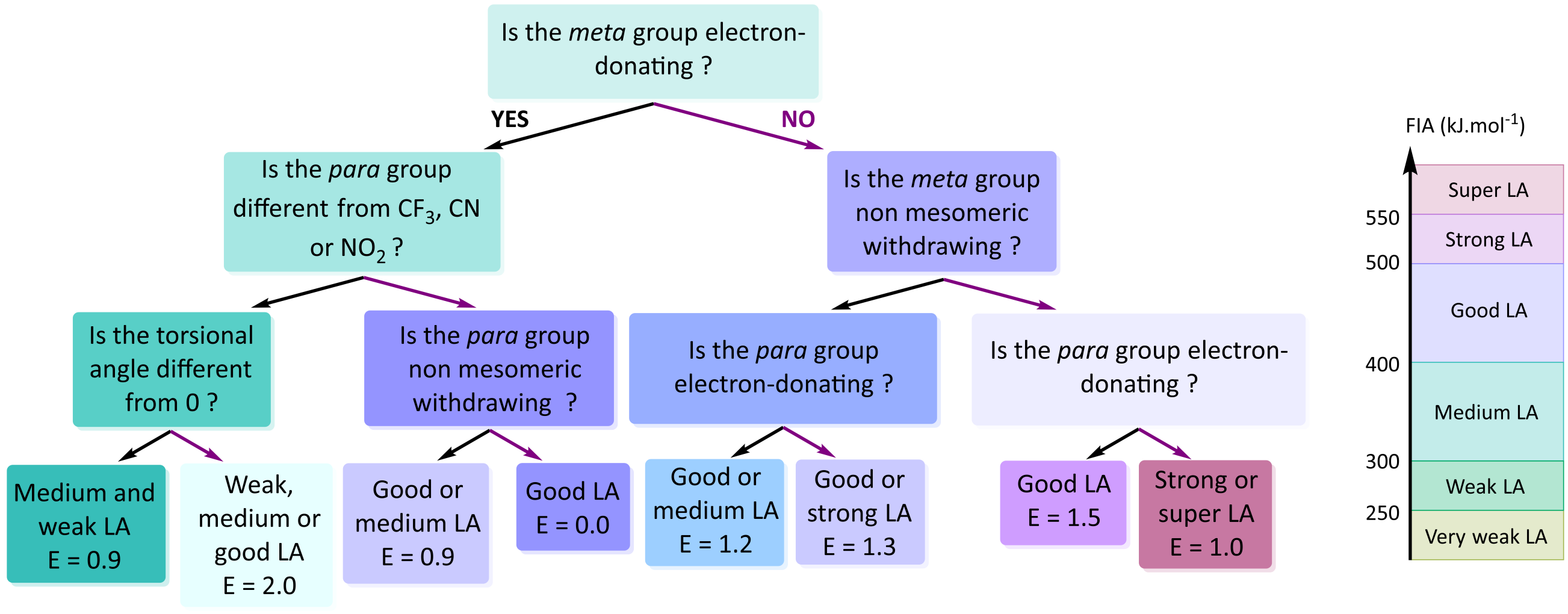

Figure S21 : Tree translation for triarylboranes (0.49 accuracy on 10-fold CV).